\documentstyle[12pt]{article}

\begin{document}

\title{\bf Causal scattering matrix in quantum electrodynamics\footnote{The extended version
of "Causal Scattering Matrix and the Chronological Product," arXiv: 1011.0859}}

\author{Yury M. Zinoviev\footnote{Electronic mail: zinoviev@mi.ras.ru}}

\date{}
\maketitle

{\it Steklov Mathematical Institute, Gubkin Street 8, 119991, Moscow,
Russia}

\vskip 0,5cm

\noindent A causal scattering matrix of quantum electrodynamics is constructed
by means of chronological product of Lagrangians where the fields have the
different arguments.  This scattering matrix is a convergent series and does not
contain the diverging integrals.

\vskip 0,5cm

\section{I. INTRODUCTION}
\setcounter{equation}{0}

The scattering matrix connects the asymptotic solutions of Schr\"odinger
equation. Stueckelberg and Rivier${}^{1}$ introduced the scattering
matrix in the paper "Causality and the $S$ matrix structure" without
making use of Schr\"odinger equation. Bogoliubov${}^{2}$ defined the
function $g(x)$ taking the values in the interval $[0,1]$ and representing
the intensity of interaction switching. Then in the space-time domains
where $g(x) = 0$ the interaction is absent, in the space-time domains
where $g(x) = 1$ it is switched on absolutely and for $0 < g(x) < 1$ it
is switched on partially. Now let $g(x)$ be not zero only in some finite
space-time domain. In this case the fields are free in the sufficiently
long ago past and in the sufficiently distant future. Bogoliubov${}^{2}$
believed that the initial and final states $\Phi (- \infty)$ and
$\Phi (\infty)$ should be connected by some operator:
$\Phi (\infty) = S(g(x))\Phi (- \infty)$. The operator $S(g(x))$ is naturally
interpreted as the scattering operator for the case when the interaction is
switched on with the intensity $g(x)$. For a small switching function
\begin{equation}
\label{3.5} S(g(x)) \approx I + i\int d^{4}x L(x) g(x).
\end{equation}
The identity operator $I$ is often omitted in Ref. 2. The operator valued
Lorentz covariant distribution $L(x)$ is the interaction Lagrangian. $L(x)$
is not decreasing at the infinity in general. The interaction operator
$L(x)g(x)$ has a compact support. The integral in the right-hand side of the
relation (\ref{3.5}) is convergent.

Poincar\'e${}^{3}$: "In the paper cited Lorentz found it necessary to supplement
his hypothesis in such a way that the relativity postulate could be valid for other
forces in addition to the electromagnetic ones. According to his idea, because of
the Lorentz transformation (and therefore because of the translational movement)
all forces behave like electromagnetic (despite their origin).

"It turned out to be necessary to consider this hypothesis more attentively
and to study the changes it makes in the gravity laws in particular. First,
it obviously enables us to suppose that the gravity forces propagate not
instantly, but at the speed of light. One could think that this is a sufficient
for rejecting such a hypothesis, because Laplace has shown that this cannot occur.
But, in fact, the effect of this propagation is largely balanced by some other
circumstance, hence, there is no any contradiction between the law proposed and
the astronomical observations.

"Is it possible to find a law satisfying the condition stated by Lorentz and
at the same time reducing to the Newton law in all the cases where the velocities
of the celestial bodies are small to neglect their squares (and also the products
of the accelerations and the distance) compared with the square of the speed of
light?"

The special relativity requires that the propagation speed
does not exceed that of light. If the propagation speed is independent
of interacting body speed, then it is equal to that of light.
The interaction force of two physical points should depend not on their
simultaneous positions and speeds but on the positions and speeds at
the time moments which differ from each other in the interval needed for
interaction force covering the distance between the physical points. The delay
is one of possible causality condition statements. The Lorentz covariance and
the causality condition are the crucial points of relativistic theory.
These conditions were proposed by Poincar\'e${}^{3}$ for the relativistic
causal gravity law. These conditions should be valid for any interaction.
In order to guarantee the theory covariance we need to demand
$$
S(g(\Lambda (A^{- 1}) x)) = U(A)S(g(x))(U(A))^{- 1},
$$
$$
\sum_{\mu, \nu \, =\, 0}^{3} \Lambda_{\nu}^{\mu} (A)
x^{\nu}\sigma^{\mu} = A\left( \sum_{\mu = 0}^{3} x^{\mu} \sigma^{\mu} \right) A^{\ast},
$$
\begin{equation}
\label{3.4}
\sigma^{0} = \left( \begin{array}{cc}

1 & 0 \\

0 & 1

\end{array} \right),
\sigma^{1} = \left( \begin{array}{cc}

0 & 1 \\

1 & 0

\end{array} \right),
\sigma^{2} = \left( \begin{array}{cc}

0 & - i \\

i &   0

\end{array} \right),
\sigma^{3} = \left( \begin{array}{cc}

1 & 0 \\

0 & - 1

\end{array} \right).
\end{equation}
$U(A)$ is an operator by means of which the free field quantum wave
functions transform under the matrix $A \in  SL(2,{\bf C})$. The second relation
(\ref{3.4}) defines the Lorentz transformation corresponding with the matrix
$A \in  SL(2,{\bf C})$. We consider the case when the space-time domain $G$ where
the function $g(x)$ is not zero is divided into two separate domains $G_{1}$ and
$G_{2}$ such that all time points of the domain $G_{1}$ lie in the past relative
to all time points of the domain $G_{2}$. (If $x^{0} < y^{0}$ in any inertial
coordinate system, the vector $x - y$ lies in the lower light cone.) Then the
function $g(x)$ may be represented as a sum of two functions
$g(x) = g_{1}(x) + g_{2}(x)$ where the function $g_{1}(x)$ is not zero in the
domain $G_{1}$ only and the function $g_{2}(x)$ is not zero in the domain $G_{2}$
only. The causality condition (Ref. 4, Section 17.5) for the scattering matrix:
\begin{equation}
\label{3.9} S(g_{1}(x) + g_{2}(x)) = S(g_{2}(x))S(g_{1}(x)).
\end{equation}
It is impossible to formulate the causality condition (\ref{3.9}) without a switching
function.

Let us consider the quantum electrodynamics with the interaction Lagrangian
(Ref. 4, relation (20.3))
\begin{equation}
\label{3.2}
L(x) = e:\bar{\psi} (x)\gamma^{\mu} \psi (x):A_{\mu} (x) =
\sum_{\alpha, \beta \, =\, 1}^{4} \sum_{\mu \, =\, 0}^{3}
e :\bar{\psi}_{\alpha}(x) \gamma_{\alpha \beta}^{\mu} \psi_{\beta}(x):
A_{\mu} (x),
\end{equation}
$$
\bar{\psi}_{\alpha} (x) = \sum_{\beta \, =\, 1}^{4}
(\psi_{\beta} (x))^{\ast} \gamma_{\beta \alpha}^{0}.
$$
$e$ is the electron charge. $4\times 4$ - matrices $\gamma_{\alpha \beta}^{\mu}$
are given by the relations (6.18) from Ref. 4. In the interaction Lagrangian (\ref{3.2})
we changed the product
$$
\bar{\psi}_{\alpha} (x) \psi_{\beta} (y)\Bigl|_{y\, =\,
x} \, \, =\, \, :\bar{\psi}_{\alpha} (x) \psi_{\beta} (x):\, \, +
\left( \sum_{\mu \, =\, 0}^{3}
\gamma_{\beta \alpha }^{\mu} \frac{\partial}{\partial x^{\mu}} +
im\right) D_{m^{2}}^{-}(x - y) \Biggl|_{y = x},
$$
$$
D_{m^{2}}^{-}(x) = \frac{i}{(2\pi)^{3}} \int d^{4}k \theta
(k^{0})\delta ((k,k) - m^{2})e^{- i(k,x)} =
$$
\begin{equation}
\label{3.8}
\frac{i}{2(2\pi)^{3}} \int d^{3}{\bf k} (|{\bf k}|^{2} + m^{2})^{- 1/2}
\exp \Biggl\{ - i(|{\bf k}|^{2} + m^{2})^{1/2}x^{0}  + i\sum_{j\, =\, 1}^{3} k^{j}x^{j} \Biggr\},
\end{equation}
$$
(k,x) = k^{0}x^{0} - \sum_{i\, =\, 1}^{3} k^{i}x^{i},
$$
for the operator valued distribution $:\bar{\psi}_{\alpha}(x) \psi_{\beta}(x):$
(Ref. 4, Section 16.2). The integral (\ref{3.8}) for $x = 0$ is divergent.

The scattering matrix $S(g(x))$ is defined by means of the $T$ - product
$T(L(x_{1})\cdots L(x_{n}))$.

\noindent Bogoliubov${}^{2}$: "Let us note as Stueckelberg did that the usual
definition of $T$ - product by means of introduction the chronological order
for the operators is effective only without the coincidence of the
arguments $x_{1},...,x_{n}$. In view of the corresponding
coefficient functions singularity their "redefinition" in the
domains of the arguments coincidence is not done explicitly and
presents a special problem...

"If we do not call attention to this difficulty and use the Wick
theorem formally, then we get the expressions of the form:
\begin{equation}
\label{3.7} \prod_{a\, <\, b} D_{m_{ab}^{2}}^{c}(x_{a} - x_{b})
\end{equation}
consisting of the causal $D^{c}$ - functions products.

"If we consider Fourier transform, then we get the integrals with the
well-known "ultra-violet" divergences."

\noindent The divergences are removed by means of renormalizations
(Ref. 4, Chapters IV, V).

\noindent Feynman (Ref. 5, Chapter 4): "The shell game that we play to
find $n$ and $j$ is technically called "renormalization". But no matter how
clever the word, it is what I would call a dippy process! Having to resort to
such hocus-pocus has prevented us from proving that the theory of quantum
electrodynamics is mathematically self-consistent. It's surprising that the
theory still hasn't been proved self-consistent one way or the other by now;
I suspect that renormalization is not mathematically legitimate. What is
certain is that we do not have a good mathematical way to describe
the theory of quantum electrodynamics: such a bunch of words to
describe the connection between $n$ and $j$ and $m$ and $e$ is not
good mathematics."

In this paper the interaction Lagrangian (\ref{3.2}) is changed for the interaction
Lagrangian $L_{3}(x_{1},x_{2},x_{3}) =$
$e:\bar{\psi} (x_{1})\gamma^{\mu} \psi (x_{2}):A_{\mu} (x_{3})$. The fields
$\bar{\psi}_{\alpha}(x_{1})$, $\psi_{\beta}(x_{2})$ and $A_{\mu}(x_{3})$ have the
different arguments. The scattering matrix is defined by means of the $T$ - product
$T(;L_{3}(x_{1},y_{1},z_{1});\cdots ;L_{3}(x_{n},y_{n},z_{n});)$. Due to
Poincar\'e${}^{3}$ the support of distributions defining this $T$ - product
lies in the closed lower light cone. The switching function $g(x)$ is changed for
the smooth switching function $h_{3}(x_{1},x_{2},x_{3})$ decreasing at the infinity.
For a small switching function $h_{3}(x_{1},x_{2},x_{3})$ the scattering matrix is
the convergent series
$$
I + ie \int d^{4}x_{1}d^{4}y_{1}d^{4}z_{1}h_{3}(x_{1},y_{1},z_{1})
:\bar{\psi} (x_{1})\gamma^{\mu} \psi (y_{1}):A_{\mu} (z_{1})\, \, +
$$
$$
\sum_{m\, =\, 2}^{\infty}  \frac{i^{m}e^{m}}{m!} \int
d^{4}x_{1}d^{4}y_{1}d^{4}z_{1}\cdots d^{4}x_{m} d^{4}y_{m}d^{4}z_{m}
h_{3}(x_{1},y_{1},z_{1})\cdots h_{3}(x_{m},y_{m},z_{m})\times
$$
\begin{equation}
\label{3.1}
T\left( ;:\bar{\psi} (x_{1})\gamma^{\mu (1)} \psi (y_{1}):A_{\mu (1)} (z_{1});\cdots ;
:\bar{\psi} (x_{m})\gamma^{\mu (m)} \psi (y_{m}):A_{\mu (m)} (z_{m});\right).
\end{equation}
The first term of the series (\ref{3.1}) coincides with the first term in the
right-hand side of the equality (\ref{3.5}). The interaction operator $L_{3}(x_{1},x_{2},x_{3})h_{3}(x_{1},x_{2},x_{3})$
should be decreasing at the infinity for the convergence of the
integral in the second term of the series (\ref{3.1}). The next terms ($m \geq 2$)
of the series (\ref{3.1}) contain the products (\ref{3.7}) for the distributions
$D^{c}(x_{a} - x_{b})$ with the different arguments  $x_{a} - x_{b}$.
These products (\ref{3.7}) are well defined and the integrals are convergent.
For a small switching function $h_{3}(x_{1},x_{2},x_{3})$ the scattering
matrix (\ref{3.1}) satisfies the relations of types (\ref{3.5}) - (\ref{3.9}).
By choosing in the series (\ref{3.1}) the switching distribution
\begin{equation}
\label{3.3} h_{3}(x_{1},x_{2},x_{3}) = g(x_{1})\delta (x_{2} - x_{1})
\delta (x_{3} - x_{1})
\end{equation}
we get the scattering matrix of Ref. 2. For the switching distribution (\ref{3.3})
the second term of the series (\ref{3.1}) coincides with the second term in the
right-hand side of the equality (\ref{3.5}) for the interaction Lagrangian (\ref{3.2}).
The next terms ($m \geq 2$) of the series (\ref{3.1}) for the switching distribution
(\ref{3.3}) contain the products for the distributions $D^{c}(x_{a} - x_{b})$ of
the same argument. For these products (\ref{3.7}) the correct definition does not
exist.

\section{II. CHRONOLOGICAL PRODUCT}
\setcounter{equation}{0}

Let us consider the complex $2\times 2$ - matrices
\begin{equation}
\label{11.1} A =  \left( \begin{array}{cc}

A_{11} & A_{12} \\

A_{21} & A_{22}

\end{array} \right), \, \,
\bar{A} = \left( \begin{array}{cc}

\bar{A}_{11} & \bar{A}_{12} \\

\bar{A}_{21} & \bar{A}_{22}

\end{array} \right), \, \,
A^{T} = \left( \begin{array}{cc}

A_{11} & A_{21} \\

A_{12} & A_{22}

\end{array} \right), \, \, A^{\ast} = (\bar{A})^{T}.
\end{equation}
If $A^{\ast} = A$, the matrix $A$ is Hermitian. The matrices $\sigma^{\mu}$
given by the relation (\ref{3.4}) form a basis of Hermitian $2\times 2$ -
matrices. The Hermitian $2\times 2$ - matrices are identified with the
Minkowski space. The complex $2\times 2$ - matrices with determinant equal to
$1$ form the group $SL(2,{\bf C})$. The complex $2\times 2$ - matrices satisfying
the equations $A^{\ast}A = \sigma^{0}$, $\det A = 1$ form the group $SU(2)$.
The group $SU(2)$ is the maximal compact subgroup of $SL(2,{\bf C})$. Let us
describe the irreducible representations of $SU(2)$. We consider the
non-negative half-integers $l = 0,1/2,1,3/2,...$. We define the
representation of $SU(2)$ on the space of polynomials with degrees $\leq 2l$
\begin{equation}
\label{1.2} T_{l}(A)\phi_{n} (z) = (A_{12}z + A_{22})^{2l}\phi_{n} \left(
 \frac{A_{11}z + A_{21}}{A_{12}z + A_{22}}\right),
\end{equation}
$$
\psi_{n} (z) = ((l - n)!(l + n)!)^{- 1/2}z^{l - n},\, \, n = - l, - l + 1,...,l - 1,l.
$$
The definition (\ref{1.2}) implies
\begin{equation}
\label{1.4} T_{l}(A)\psi_{n} (z) =
\sum_{m\, =\, - \, l,\, - \, l\, +\, 1,\, ...,\, l\, - \,1,\, l}
\psi_{m} (z)t_{mn}^{l}(A),
\end{equation}
\begin{eqnarray}
\label{1.5} t_{mn}^{l}(A) = ((l - m)!(l + m)!(l - n)!(l + n)!)^{1/2}
\times \nonumber \\ \sum_{j\, =\, -\, \infty}^{\infty} \frac{A_{11}^{l - m
- j}A_{12}^{j}A_{21}^{m - n + j}A_{22}^{l + n - j}}{\Gamma (j +
1)\Gamma (l - m - j + 1)\Gamma (m - n + j + 1)\Gamma (l + n - j +
1)}
\end{eqnarray}
where $\Gamma (z)$ is the gamma - function. The function $(\Gamma
(z))^{- 1}$ equals zero for $z = 0 , - 1, - 2,...$. Therefore the
series (\ref{1.5}) is the polynomial. The relation (\ref{1.2}) defines
the representation of the group $SU(2)$. Thus the polynomial (\ref{1.5})
defines the representation of $SU(2)$. This $(2l + 1)$ -
dimensional representation is irreducible. The relation (\ref{1.5})
has an analytic continuation to the matrices $A \in  SL(2,{\bf C})$.

Let us consider the free real scalar field $\varphi (x)$, the free electromagnetic
field $A_{\mu}(x)$ and the free spin field $\psi_{\alpha}(x)$. The operator
valued distributions $\varphi (x)$, $A_{\mu}(x)$, $\psi_{\alpha}(x)$ take the
values in the set of Hilbert space operators. The commutation and anticommutation
relations (11.3), (12.4), (13.4) from Ref. 4 are
$$
[\varphi (x),\varphi (y)] = - i D_{m^{2}}(x - y) =
- i \left( D_{m^{2}}^{-}(x - y) - D_{m^{2}}^{-}(y - x) \right),
$$
$$
[A_{\mu}(x),A_{\nu}(y)] = i\eta^{\mu \nu} D_{0}(x - y),
$$
\begin{equation}
\label{2.1}
\psi_{\alpha} (x)\bar{\psi}_{\beta} (y) + \bar{\psi}_{\beta} (y) \psi_{\alpha} (x)
= \left( \sum_{\mu \, =\, 0}^{3} \gamma_{\alpha
\beta}^{\mu} \frac{\partial}{\partial x^{\mu}} - im\right)
D_{m^{2}}(x - y).
\end{equation}
The diagonal Minkowski $4\times 4$ - matrix $\eta^{\mu \nu}$ has the
diagonal matrix elements $\eta^{00} = - \eta^{11} = - \eta^{22} = - \eta^{33} = 1$.
The Pauli - Jordan distribution $D_{m^{2}}(x)$ is given by the relation (10.18)
from Ref. 4 (relations (\ref{3.8}), (\ref{2.1})). The commutation relation for the
free vector field $U_{\mu}(x)$ is given by the relation (11.27) from Ref. 4. Th1s
relation is similar to the relations (\ref{2.1}). The commutation relations for
another free field pairs are equal to zero. The free spin field $\psi_{\alpha}(x)$
commutes with the free fields $\varphi (x)$, $U_{\mu}(x)$, $A_{\mu}(x)$ in particular.
The operator valued distributions $\varphi (x)$, $U_{\mu}(x)$, $A_{\mu}(x)$,
$\psi_{\alpha}(x)$ and all its possible derivatives are called the free quantum fields
and denote $u_{\alpha}(x)$. Due to the relations (2.31), (2.33) from Ref. 6 the free
field quantum wave functions $\psi_{m \dot{m}} (x)$ transform under the matrix
$A \in  SL(2,{\bf C})$ as
$$
(U(A)\psi)_{m \dot{m}} (x) =
\sum_{n\, =\, 1}^{2l\, +\, 1} \sum_{\dot{n} \, =\, 1}^{2\dot{l} \, +\, 1}
t_{m\, -\, l\, -\, 1,\, n\, -\, l\, -\, 1}^{l}(A)
t_{\dot{m} \, -\, \dot{l} \, -\, 1,\, \dot{n} \, -\, \dot{l} \, -\,
1}^{\dot{l}}(\bar{A}) \psi_{n\dot{n}} \left( \Lambda (A^{- 1}) x \right)  +
$$
\begin{equation}
\label{2.101}
\sum_{n\, =\, 2l\, +\, 2}^{4l\, +\, 2} \sum_{\dot{n} \, =\, 1}^{2\dot{l} \, +\, 1}
t_{m\, -\, 3l\, -\, 2,\, n\, -\, 3l\, -\,
2}^{l}(((\bar{A})^{T})^{- 1})t_{\dot{m} \, -\, \dot{l} \, -\, 1,\, \dot{n} \, -\, \dot{l} \, -\,
1}^{\dot{l}}((A^{T})^{- 1}) \psi_{n\dot{n}} \left( \Lambda (A^{- 1}) x \right).
\end{equation}
The representation (\ref{2.101}) is reducible. The free field quantum wave functions
$\psi_{m \dot{m}} (x)$ may transform under $T_{l}(A)\times T_{\dot{l}}(\bar{A})$
representation of $SL(2,{\bf C})$: the wave functions $\psi_{m \dot{m}} (x) = 0$
for the indices $m = 2l + 2,...,4l + 2$, $\dot{m} = 1,...,2\dot{l} + 1$.
The free field quantum wave functions $\psi_{m \dot{m}} (x)$ may transform under
$T_{l}(((\bar{A})^{T})^{- 1})\times T_{\dot{l}}((A^{T})^{- 1})$
representation of $SL(2,{\bf C})$: the wave functions $\psi_{m \dot{m}} (x) = 0$
for the indices $m = 1,...,2l + 1$, $\dot{m} = 1,...,2\dot{l} + 1$.
The free field $u_{\alpha}(x)$ transforms under the matrix $A \in  SL(2,{\bf C})$
similar to the right-hand side of the relation (\ref{2.101}). For the free Fermi-field
$u_{\alpha}(x)$ the integer $2l + 2\dot{l}$ is odd. The free spin field $\psi_{\alpha}(x)$
transforms under the $SL(2,{\bf C})$ representation (\ref{2.101}) for $l = 1/2$,
$\dot{l} = 0$ and the integer $2l + 2\dot{l} = 1$. The adjoint free spin field
$(\psi_{\alpha} (x))^{\ast}$ transforms under the $SL(2,{\bf C})$ representation
(\ref{2.101}) for $l = 0$, $\dot{l} = 1/2$ and the integer $2l + 2\dot{l} = 1$. For
the free Fermi-fields the Klein-Gordon equation implies the equation of Dirac type
(see Ref. 6). For the free Bose-field $u_{\alpha}(x)$ the integer $2l + 2\dot{l}$ is
even. The free real scalar field $\varphi (x)$ transforms under $T_{0}(A)\times T_{0}(\bar{A})$
representation of $SL(2,{\bf C})$ and the integer $2l + 2\dot{l} = 0$. The free vector
field $U_{\mu}(x)$ and the free electromagnetic field $A_{\mu}(x)$ transform under
$T_{1}(A)\times T_{0}(\bar{A})$ representation of $SL(2,{\bf C})$ and the integer
$2l + 2\dot{l} = 2$.

Let $\Omega$ be the vacuum vector of Hilbert space. The vacuum expectations of
the products of two free fields are given by the relations (10.17), (16.12) - (16.14)
from Ref. 4
$$
(\Omega, \varphi (x)\varphi (y)\Omega) \, \, =\, \,
<\varphi (x)\varphi (y)>_{0} \, \, =
\, \, - i D_{m^{2}}^{-}(x - y),
$$
$$
(\Omega, A_{\mu}(x)A_{\nu}(y)\Omega) \, \, = \, \,
<A_{\mu}(x)A_{\nu}(y)>_{0}\, \, =\, \, i\eta^{\mu \nu} D_{0}^{-}(x - y),
$$
\begin{equation}
\label{2.3}
(\Omega, \psi_{\alpha} (x)\bar{\psi}_{\beta} (y)\Omega) \, \, = \, \,
<\psi_{\alpha} (x)\bar{\psi}_{\beta} (y)>_{0} \, \, =\, \, \left(
\sum_{\mu \, =\, 0}^{3} \gamma_{\alpha \beta}^{\mu}
\frac{\partial}{\partial x^{\mu}} - im\right) D_{m^{2}}^{-}(x - y).
\end{equation}
Let us assume $<I>_{0}\, \, = 1$. Then the relations (\ref{2.1}), (\ref{2.3})
imply the vacuum expectation
\begin{equation}
\label{3.55} (\Omega, \bar{\psi}_{\alpha} (x)\psi_{\beta} (y) \Omega) \, \, =\, \,
<\bar{\psi}_{\alpha} (x)\psi_{\beta} (y)>_{0} \, \, =\, \,
\left(  \sum_{\mu \, =\, 0}^{3} \gamma_{\beta \alpha} ^{\mu}
\frac{\partial}{\partial x^{\mu}} + im\right) D_{m^{2}}^{-}(x - y).
\end{equation}
The vacuum expectation $<U_{\lambda}^{\ast}(x)U_{\nu}(y)>_{0}$ is similar to
the vacuum expectations (\ref{2.3}). The vacuum expectations of another free
field products are either the derivatives of the distributions (\ref{2.3}),
(\ref{3.55}) or are equal to zero.

The free field normal product is given by the relations (16.17) from Ref. 4:
$$
:I:\, = I,\, \, :u_{\alpha} (x):\, = u_{\alpha} (x),
$$
$$
u_{\alpha (1)} (x_{1}) \cdots u_{\alpha (n)} (x_{n}) =\, \,
:u_{\alpha (1)} (x_{1}) \cdots u_{\alpha (n)} (x_{n}):\, \, +
$$
$$
\sum_{1\, \leq \, k\, <\, l\, \leq \, n} \, \, (- 1)^{\sigma (k,\,
l,1,...,\widehat{k},...,\widehat{l},...,n)}
<u_{\alpha (k)} (x_{k}) u_{\alpha (l)} (x_{l})>_{0} \times
$$
\begin{equation}
\label{2.4} :u_{\alpha (1)} (x_{1}) \cdots \widehat{u_{\alpha (k)} (x_{k})}
\cdots \widehat{u_{\alpha (l)} (x_{l})} \cdots u_{\alpha (n)}
(x_{n}): + \cdots,\, \,  n = 2,3,..
\end{equation}
The subsequent summings in the right-hand side of the last
equality (\ref{2.4}) run over two pairs of numbers from $1,...,n$, run over
three pairs of numbers from $1,...,n$, etc. The permutation of Fermi-operators
corresponding with the permutation $1,2,...,n \rightarrow j(1),j(2),...,j(n)$
has the parity
\begin{equation}
\label{2.701} \sigma (j(1),j(2),...,j(n)) = \sum_{k\, =\, 1}^{n}
\, \, \sum_{1\, \leq \, i\, <\, j(k), \, \, i\, \neq \,
j(1),...,j(k)} (2l_{j(k)} + 2\dot{l}_{j(k)}) (2l_{i} +
2\dot{l}_{i}) \, \, \hbox{mod} \, 2.
\end{equation}
(If the integer $i$ satisfying the inequalities $1 \leq  i < j(k)$,
$i \neq j(1),...,j(k)$ does not exist, then we assume
$2l_{i} + 2\dot{l}_{i} = 0 \, \, \hbox{mod} \, 2$.) The definition
(\ref{2.4}) is called the Wick theorem for the normal products in
Section 16.2 from Ref. 4. The explicit definition (\ref{2.701}) is
absent. The numbers (\ref{2.701}) are assumed to be equal to zero in
"Causal Scattering Matrix and the Chronological Product," arXiv: 1011.0859.
It is correct for the free Bose-fields only. The definition (\ref{2.701})
implies for $n = 2$
\begin{equation}
\label{2.700} \sigma (1,2) = 0,\, \, \sigma (2,1) = (2l_{1} +
2\dot{l}_{1}) (2l_{2} + 2\dot{l}_{2}) \, \, \hbox{mod} \, 2
\end{equation}
and for $n = 3$
$$
\sigma (1,2,3) = 0,\, \, \sigma (2,1,3) = (2l_{1} + 2\dot{l}_{1})
(2l_{2} + 2\dot{l}_{2}) \, \, \hbox{mod} \, 2,
$$
$$
\sigma (1,3,2) = (2l_{2} + 2\dot{l}_{2}) (2l_{3} + 2\dot{l}_{3}) \,
\, \hbox{mod} \, 2,
$$
$$
\sigma (3,1,2) = (2l_{1} + 2\dot{l}_{1} + 2l_{2} + 2\dot{l}_{2})
(2l_{3} + 2\dot{l}_{3}) \, \, \hbox{mod} \, 2,
$$
$$
\sigma (2,3,1) = (2l_{1} + 2\dot{l}_{1}) (2l_{2} + 2\dot{l}_{2} +
2l_{3} + 2\dot{l}_{3})\, \, \hbox{mod} \, 2,
$$
\begin{equation}
\label{2.702} \sigma (3,2,1) = (2l_{1} + 2\dot{l}_{1}) (2l_{2} +
2\dot{l}_{2}) + (2l_{1} + 2\dot{l}_{1} + 2l_{2} + 2\dot{l}_{2})
(2l_{3} + 2\dot{l}_{3}) \, \, \hbox{mod} \, 2.
\end{equation}
Due to the first relation (\ref{2.700}) the relation (\ref{2.4}) for $n = 2$ is
\begin{eqnarray}
\label{2.402} u_{\alpha (1)} (x_{1}) u_{\alpha (2)} (x_{2}) =\, \,
:u_{\alpha (1)} (x_{1}) u_{\alpha (2)} (x_{2}): + <u_{\alpha (1)}
(x_{1}) u_{\alpha (2)} (x_{2})>_{0}.
\end{eqnarray}
The normal product may be also defined in the following way
$$
:u_{\alpha (1)} (x_{1}) \cdots u_{\alpha (n)} (x_{n}): \, \, =
u_{\alpha (1)} (x_{1}) \cdots u_{\alpha (n)} (x_{n})\, \, - \sum_{1\,
\leq \, k\, <\, l\, \leq \, n} (- 1)^{\sigma (k,\,
l,1,...,\widehat{k},...,\widehat{l},...,n)} \times
$$
\begin{equation}
\label{2.5} <u_{\alpha (k)} (x_{k}) u_{\alpha (l)} (x_{l})>_{0}
u_{\alpha (1)} (x_{1}) \cdots \widehat{u_{\alpha (k)} (x_{k})}
\cdots \widehat{u_{\alpha (l)} (x_{l})} \cdots u_{\alpha (n)}
(x_{n}) + \cdots,
\end{equation}
$n = 2,3$,.. The subsequent summings run over two pairs of numbers from
$1,..,n$, run over three pairs of numbers from $1,..,n$, etc. The summing over
the even (odd) number of pairs has the sign plus (minus). The relation
(\ref{2.5}) for $n = 2$ coincides with the relation (\ref{2.402}).
Let us prove the relation (\ref{2.5}) by making use of the relation
(\ref{2.4}). Let us change every distribution $<u_{\alpha (k)}
(x_{k}) u_{\alpha (l)} (x_{l})>_{0}$ in the right-hand side of the
relation (\ref{2.4}) for
\begin{equation}
\label{2.51} <u_{\alpha (k)} (x_{k}) u_{\alpha (l)} (x_{l})>_{0}  -
<u_{\alpha (k)} (x_{k}) u_{\alpha (l)} (x_{l})>_{0}.
\end{equation}
The distribution (\ref{2.51}) is equal to zero. We get the relation
\begin{eqnarray}
\label{2.52} :u_{\alpha (1)} (x_{1}) \cdots u_{\alpha (n)}
(x_{n}):\, \, = \, \, :u_{\alpha (1)} (x_{1}) \cdots u_{\alpha (n)}
(x_{n}): + \nonumber
\\ \sum_{1\, \leq \, k\, <\, l\, \leq \, n} (<u_{\alpha (k)} (x_{k})
u_{\alpha (l)} (x_{l})>_{0}  -
<u_{\alpha (k)} (x_{k}) u_{\alpha (l)} (x_{l})>_{0}) \times \nonumber \\
(- 1)^{\sigma (k,\, l,1,...,\widehat{k},...,\widehat{l},...,n)}
:u_{\alpha (1)} (x_{1}) \cdots \widehat{u_{\alpha (k)} (x_{k})}
\cdots \widehat{u_{\alpha (l)} (x_{l})} \cdots u_{\alpha (n)}
(x_{n}): + \cdots
\end{eqnarray}
Choose the first term $<u_{\alpha (k)} (x_{k}) u_{\alpha (l)} (x_{l})>_{0}$
in every sum (\ref{2.51}) of the equality (\ref{2.52})
$$
<u_{\alpha (k)} (x_{k}) u_{\alpha (l)} (x_{l})>_{0} (- 1)^{\sigma
(k,\, l,1,...,\widehat{k},...,\widehat{l},...,n)}\times
$$
$$
:u_{\alpha (1)} (x_{1}) \cdots \widehat{u_{\alpha (k)} (x_{k})}
\cdots \widehat{u_{\alpha (l)} (x_{l})} \cdots u_{\alpha (n)}
(x_{n}):,...
$$
Adding the first term $:u_{\alpha (1)} (x_{1}) \cdots u_{\alpha (n)} (x_{n}):$
to the above terms we get the first term
$u_{\alpha (1)} (x_{1}) \cdots u_{\alpha (n)} (x_{n})$ of the right-hand side
of the relation (\ref{2.5}) due to the relation (\ref{2.4}). Let us choose the
second term $ - <u_{\alpha (k)} (x_{k}) u_{\alpha (l)} (x_{l})>_{0}$ in one sum
(\ref{2.51}) of the relation (\ref{2.52}) and choose the first term
$<u_{\alpha (k)} (x_{k}) u_{\alpha (l)} (x_{l})>_{0}$ in all the other sums
(\ref{2.51}). Due to the relation (\ref{2.4}) we get the second term of the
right-hand side of the relation (\ref{2.5}). If we continue this process, we
transform the relation (\ref{2.52}) into the relation (\ref{2.5}).

By making use of the definition (\ref{2.5}) it is possible to prove the
following relation
\begin{equation}
\label{2.53}  :u_{\alpha (j(1))}(x_{j(1)}) \cdots u_{\alpha
(j(n))}(x_{j(n)}):\, \,  = (- 1)^{\sigma (j(1),j(2),...,j(n))}
:u_{\alpha (1)}(x_{1}) \cdots u_{\alpha (n)}(x_{n}):
\end{equation}
for any permutation $1,...,n \rightarrow j(1),j(2),...,j(n)$.
The relations (\ref{2.1}), (\ref{2.402}) imply the relation
(\ref{2.53}) for $n = 2$. The relation (\ref{2.5}) may be rewritten
in two recurrent ways
$$
:u_{\alpha (1)} (x_{1}) u_{\alpha (2)} (x_{2}) \cdots u_{\alpha (n)}
(x_{n}):\, \,  =
$$
$$
u_{\alpha (1)} (x_{1})\, \, :u_{\alpha (2)} (x_{2}) \cdots u_{\alpha
(n)} (x_{n}):\, \, -
$$
$$
\sum_{2\, \leq \, l\, \leq \, n} (- 1)^{\sigma (1,\,
l,2,...,\widehat{l},...,n)}  <u_{\alpha (1)} (x_{1}) u_{\alpha (l)}
(x_{l})>_{0} \times
$$
\begin{equation}
\label{2.541}
:u_{\alpha (2)} (x_{2}) \cdots \widehat{u_{\alpha (l)} (x_{l})}
\cdots u_{\alpha (n)} (x_{n}):,
\end{equation}
$$
:u_{\alpha (1)} (x_{1}) \cdots u_{\alpha (n - 1)} (x_{n - 1})
u_{\alpha (n)} (x_{n}):\, \,  =
$$
$$
:u_{\alpha (1)} (x_{1}) \cdots u_{\alpha (n - 1)} (x_{n - 1}):\, \,
u_{\alpha (n)} (x_{n})  -
$$
$$
\sum_{1\, \leq \, k\, <\, n} (- 1)^{\sigma (k,\,
n,1,...,\widehat{k},...,n - 1)}  <u_{\alpha (k)} (x_{k}) u_{\alpha
(n)} (x_{n})>_{0} \times
$$
\begin{equation}
\label{2.542} :u_{\alpha (1)} (x_{1}) \cdots \widehat{u_{\alpha (k)}
(x_{k})} \cdots u_{\alpha (n - 1)} (x_{n - 1}):,\, \, n = 2,3,...
\end{equation}
By making use of the relations (\ref{2.541}), (\ref{2.542}) we prove
the relations
$$
:u_{\alpha (1)} (x_{1})u_{\alpha (3)} (x_{3})u_{\alpha (2)}
(x_{2}):\, \, = (- 1)^{(2l_{2} + 2\dot{l}_{2}) (2l_{3} +
2\dot{l}_{3})}\, \, :u_{\alpha (1)} (x_{1})u_{\alpha (2)}
(x_{2})u_{\alpha (3)} (x_{3}):,
$$
\begin{equation}
\label{2.55} :u_{\alpha (2)} (x_{2})u_{\alpha (1)} (x_{1})u_{\alpha
(3)} (x_{3}):\, \, = (- 1)^{(2l_{1} + 2\dot{l}_{1}) (2l_{2} +
2\dot{l}_{2})}\, \, :u_{\alpha (1)} (x_{1})u_{\alpha (2)}
(x_{2})u_{\alpha (3)} (x_{3}):
\end{equation}
and the relation (\ref{2.53}) for an arbitrary $n$.

The chronological product of field operators is defined by
\begin{equation}
\label{2.7} T(u_{\alpha (1)}(x_{1}) \cdots u_{\alpha (n)}(x_{n})) =
(- 1)^{\sigma (j(1),j(2),...,j(n))} u_{\alpha (j(1))}(x_{j(1)})
\cdots u_{\alpha (j(n))}(x_{j(n)}),
\end{equation}
$$
x_{j(1)}^{0} > x_{j(2)}^{0} > \cdots > x_{j(n)}^{0}.
$$
The chronological product
$T(u_{\alpha (1)}(x_{1}) \cdots u_{\alpha (n)}(x_{n}))$ in Ref. 4 is
defined by the relation (19.1) for the closed set
$x_{j_{1}}^{0} \geq x_{j_{2}}^{0} \geq \cdots \geq x_{j_{n}}^{0}$
instead of the open set
$x_{j_{1}}^{0} > x_{j_{2}}^{0} > \cdots > x_{j_{n}}^{0}$ in the
definition (\ref{2.7}). The distribution may be restricted only
to an open set. Bogoliubov${}^{2}$: "Let us note as Stueckelberg
did that the usual definition of $T$ - product by means of introduction
the chronological order for operators is effective only without the
coincidence of the arguments $x_{1},...,x_{n}$. In view of the
corresponding coefficient functions singularity their "redefinition"
in the domains of the arguments coincidence is not done explicitly
and presents a special problem."  The correct relation (\ref{2.7}) does
not define the chronological product in the domains of the time arguments
coincidence.

We give the additional definition of chronological product for two free
field operators
\begin{equation}
\label{2.18} T(u_{\alpha (1)}(x_{1})u_{\alpha (2)}(x_{2})) =
u_{\alpha (1)}(x_{1})u_{\alpha (2)}(x_{2}) + <u_{\alpha
(1)}(x_{1})u_{\alpha (2)}(x_{2})>_{c},
\end{equation}
$$
<u_{\alpha (1)}(x_{1})u_{\alpha (2)}(x_{2}) >_{c} =
f_{\alpha (1), \alpha (2)} (x_{1} - x_{2}),\, \,
f_{\alpha (1), \alpha (2)} (x) \in S^{\prime}({\bf R}^{4}).
$$
The index in the average $<>_{c}$ means causal. Poincar\'e${}^{3}$ causality
condition: the  support of $f_{\alpha (1), \alpha (2)} (x)$ lies in the closed
lower light cone. The chronological product
$T(u_{\alpha (1)}(x_{1});u_{\alpha (2)}(x_{2}))$
differs from the usual product
$u_{\alpha (1)}(x_{1})u_{\alpha (2)}(x_{2})$ in the only case when
the argument difference $x_{1} - x_{2}$ lies in the closed lower light cone.
The relation (\ref{2.7}) implies
\begin{equation}
\label{2.22} <T(u_{\alpha (1)}(x_{1})u_{\alpha (2)}(x_{2}))>_{0} \, \, =
\left\{ {<u_{\alpha (1)}(x_{1})u_{\alpha
(2)}(x_{2})>_{0}, \hskip 2cm x_{1}^{0} > x_{2}^{0},} \atop {(-
1)^{\sigma (2,1)} <u_{\alpha
(2)}(x_{2})u_{\alpha (1)}(x_{1})>_{0}, \hskip 0,3cm x_{1}^{0} <
x_{2}^{0},} \right.
\end{equation}
The number $\sigma (2,1)$ is given by the second relation (\ref{2.700}).
We assume $<I>_{0}\, \, = 1$. Then the definition (\ref{2.18})
implies
\begin{equation}
\label{2.23} <T(u_{\alpha (1)}(x_{1})u_{\alpha (2)}(x_{2}))>_{0} \, \, =
\, \, <u_{\alpha (1)}(x_{1})u_{\alpha (2)}(x_{2})>_{0} + <u_{\alpha
(1)}(x_{1})u_{\alpha (2)}(x_{2})>_{c}.
\end{equation}
Due to the second relation (\ref{3.8}) the distribution $D_{m^{2}}^{-}(x)$ satisfies
the Klein - Gordon equation
\begin{equation}
\label{2.10} ((\partial_{x}, \partial_{x}) + m^{2})D_{m^{2}}^{-}(x)
= 0, \, \, (\partial_{x}, \partial_{x}) =
\left( \frac{\partial}{\partial x^{0}}\right)^{2} - \sum_{i \, =\, 1}^{3}
\left( \frac{\partial}{\partial x^{i}}\right)^{2}.
\end{equation}
The relations (\ref{2.3}), (\ref{2.22}) - (\ref{2.10}) imply
\begin{equation}
\label{2.24} ((\partial_{x_{1}}, \partial_{x_{1}}) +
m^{2})<u_{\alpha (1)}(x_{1})u_{\alpha (2)}(x_{2})>_{c}\, \, =\, \,
((\partial_{x_{1}}, \partial_{x_{1}}) + m^{2})f_{\alpha (1),
\alpha (2)} (x_{1} - x_{2})
= 0,
\end{equation}
$x_{1}^{0} \neq x_{2}^{0}$. The support of
$f_{\alpha (1), \alpha (2)} (x) \in S^{\prime}({\bf R}^{4})$
lies in the closed lower light cone. It is possible
therefore to continue the relation (\ref{2.24}) into the domain
$x_{1} \neq x_{2}$. The general form of the distribution (\ref{2.24})
with support at the point $x = 0$ is
\begin{equation}
\label{2.25} ((\partial_{x}, \partial_{x}) +
m^{2}) f_{\alpha (1), \alpha (2)} (x) =
P(\partial_{x}) \delta (x)
\end{equation}
where $P(x)$ is a polynomial. The Klein - Gordon fundamental equation
\begin{equation}
\label{2.11} ((\partial_{x}, \partial_{x}) + m^{2}) e_{m^{2}}(x)
= \delta (x)
\end{equation}
has the unique solution in the class of distributions with supports in
the closed upper light cone. Let the equation (\ref{2.11}) have two solutions
$e^{(1)}(x)$, $e^{(2)}(x)$ with supports  in the closed upper light cone.
Since the supports of $e^{(1)}(x)$, $e^{(2)}(x)$ lie in the closed upper
light cone, the convolution is defined. The convolution commutativity
$$
\int  d^{4}x d^{4}ye^{(1)}(x - y)e^{(2)}(y) \phi (x) =
\int d^{4}x d^{4}ye^{(2)}(x - y)e^{(1)}(y) \phi (x) =
$$
\begin{equation}
\label{2.26}
\int  d^{4}x d^{4}ye^{(1)}(x)e^{(2)}(y) \phi (x + y)
\end{equation}
implies the coincidence of these solutions
\begin{eqnarray}
\label{2.12} e^{(2)}(x) = ((\partial_{x}, \partial_{x}) + m^{2})
\int d^{4}ye^{(1)}(x - y)e^{(2)}(y) = \nonumber \\ ((\partial_{x},
\partial_{x}) + m^{2}) \int d^{4}ye^{(2)}(x - y)e^{(1)}(y) =
e^{(1)}(x).
\end{eqnarray}
Therefore the solution $e_{m^{2}}(x)$ of the equation (\ref{2.11}) coincides
with the distribution $D_{m^{2}}^{ret}(x)$ given by the relation (14.7) from Ref. 4
\begin{equation}
\label{2.13} e_{m^{2}}(x) = D_{m^{2}}^{ret}(x) = \lim_{\epsilon \, \rightarrow \,
+\, 0} \frac{1}{(2\pi)^{4}} \int d^{4}k \frac{e^{- i(k,x)}}{m^{2} -
(k^{0} + i\epsilon)^{2} + |{\bf k}|^{2}}.
\end{equation}
The distribution $<u_{\alpha (1)}(x_{1})u_{\alpha (2)}(x_{2})>_{c}$
has the form $f_{\alpha (1), \alpha (2)} (x_{1} - x_{2})$. The distribution
$f_{\alpha (1), \alpha (2)} (x) \in S^{\prime}({\bf R}^{4})$
has the support in the closed lower light cone. Hence the equation (\ref{2.25})
solution is
\begin{equation}
\label{2.27} <u_{\alpha (1)}(x_{1})u_{\alpha (2)}(x_{2})>_{c}
= P(\partial_{x_{1}}) D_{m^{2}}^{ret} (x_{2} - x_{1}).
\end{equation}
Due to the relation (14.8) from Ref. 4
\begin{equation}
\label{2.21} D_{m^{2}}^{ret}(x) = \left\{ {D_{m^{2}}(x), \hskip
0,5cm x^{0}
> 0,} \atop  {0, \hskip 1,7cm x^{0} < 0.}\right.
\end{equation}
The relations (\ref{2.1}), (\ref{2.21}) are crucial for the chronological product definition
(\ref{2.7}), (\ref{2.18}). Let us suppose
$$
<\varphi(x)\varphi(y)>_{c} = - i D_{m^{2}}^{ret}(y - x),\, \,
<A_{\mu}(x)A_{\nu}(y)>_{c} = i\eta^{\mu \nu} D_{0}^{ret}(y - x),
$$
$$
<\psi_{\alpha} (x)\bar{\psi_{\beta}} (y)>_{c} \, \, =\, \,
\left(  \sum_{\mu \, =\, 0}^{3} \gamma_{\alpha \beta} ^{\mu}
\frac{\partial}{\partial x^{\mu}} - im\right)
D_{m^{2}}^{ret}(y - x),
$$
\begin{equation}
\label{2.17}
<\bar{\psi}_{\alpha} (x)\psi_{\beta} (y)>_{c} \, \, =\, \,
\left(  \sum_{\mu \, =\, 0}^{3} \gamma_{\beta \alpha} ^{\mu}
\frac{\partial}{\partial x^{\mu}} + im\right)
D_{m^{2}}^{ret}(y - x).
\end{equation}
The distribution $<U_{\lambda}^{\ast}(x)U_{\nu}(y)>_{c}$ is
similar to the distributions (\ref{2.17}). The distributions
$<u_{\alpha (1)}(x_{1})u_{\alpha (2)}(x_{2})>_{c}$ for other free
fields are the derivatives of the distributions (\ref{2.17}) or are
equal to zero. The relations (\ref{2.1}), (\ref{2.18}), (\ref{2.21}),
(\ref{2.17}) imply the relation (\ref{2.7}) for $n = 2$. The relations
(\ref{2.18}), (\ref{2.17}) yield the complete definition for the chronological
product of two free field operators. The distribution
$<u_{\alpha (1)}(x_{1})u_{\alpha (2)} (x_{2})>_{c}$ gives the delay.
For $m = 0$ the distribution (\ref{2.13}) has the form (Ref. 7, Section 30)
\begin{equation}
\label{2.29}
D_{0}^{ret}(- x) = (2\pi)^{- 1} \theta (- x^{0})\delta ((x,x)).
\end{equation}
The distribution (\ref{2.29}) is the Lorentz invariant formulation  of delay.

Let us calculate the distribution (\ref{2.23}). The distribution
\begin{equation}
\label{2.28}
D_{m^{2}}^{c}(x) \equiv \lim_{\epsilon \, \rightarrow \, + 0}
\frac{1}{(2\pi)^{4}} \int d^{4}k \frac{e^{i(k,x)}}{m^{2} -
(k,k) - i\epsilon}.
\end{equation}
is defined by the relation (14.13) from Ref. 4. Due to the
relation (14.12) from Ref. 4
\begin{equation}
\label{2.91} D_{m^{2}}^{c}(x) = \left\{ {D_{m^{2}}^{-}(x), \hskip
0,5cm x^{0}
> 0,} \atop  {D_{m^{2}}^{-}(- x), \hskip 0,3cm x^{0} < 0.}\right.
\end{equation}
Hence the distribution $D_{m^{2}}^{c}(x) - D_{m^{2}}^{-}(x) = 0$ for
$x^{0} > 0$. In view of the definitions (\ref{3.8}), (\ref{2.28}) the
distribution $D_{m^{2}}^{c}(x) - D_{m^{2}}^{-}(x)$ is Lorentz invariant.
Hence the support of this distribution lies in the closed lower light
cone.  The distribution $D_{m^{2}}^{c}(x)$ is the solution of the
equation (\ref{2.11}). Hence the distribution
$D_{m^{2}}^{c}(x) - D_{m^{2}}^{-}(x)$ satisfies the equation (\ref{2.11})
and  coincides with the distribution $D_{m^{2}}^{ret}(- x)$
\begin{equation}
\label{2.15} D_{m^{2}}^{c}(x) = D_{m^{2}}^{ret}(- x) +
D_{m^{2}}^{-}(x).
\end{equation}
The change $x \rightarrow - x$ in the relation (\ref{2.15}) yields the relation
proved in Section 14.2 from Ref. 4. The equalities $<1>_{0}\, \, =\, \, 1$ and
(\ref{2.3}), (\ref{3.55}), (\ref{2.23}), (\ref{2.17}), (\ref{2.15}) imply
$$
<T(\varphi (x)\varphi (y))>_{0} = - i D_{m^{2}}^{c}(x - y),\, \,
<T(A_{\lambda}(x)A_{\nu}(y))>_{0} =  i\eta^{\mu \nu} D_{0}^{c}(x - y),
$$
$$
<T(\psi_{\alpha} (x)\bar{\psi}_{\beta} (y))>_{0} \, \, =\, \,
\left( \sum_{\mu \, =\, 0}^{3} \gamma_{\alpha
\beta}^{\mu} \frac{\partial}{\partial x^{\mu}} - im\right)
D_{m^{2}}^{c}(x - y),
$$
\begin{equation}
\label{2.9} <T(\bar{\psi}_{\alpha} (x)\psi_{\beta} (y))>_{0} \, \, =\, \,
\left( \sum_{\mu \, =\, 0}^{3} \gamma_{\beta \alpha}^{\mu}
\frac{\partial}{\partial x^{\mu}} + im\right) D_{m^{2}}^{c}(x - y).
\end{equation}
The vacuum expectation $<T(U_{\lambda}^{\ast}(x);U_{\nu}(y))>_{0}$ is
similar to the vacuum expectations (\ref{2.9}). The distributions $<T(u_{\alpha
(1)}(x_{1})u_{\alpha (2)}(x_{2}))>_{0}$ for other free fields are
the derivatives of the distributions (\ref{2.9}) or are equal to zero.
The distribution $D_{m^{2}}^{c}(x - y)$ defines the vacuum
expectation of the chronological product (\ref{2.18})
of two free quantum fields. Stueckelberg and Rivier${}^{1}$:
the classical "causal action" is given by the distribution
$D_{m^{2}}^{ret}(y - x)$ and the distribution $D_{m^{2}}^{c}(x - y)
= D_{m^{2}}^{c}(y - x)$ defines the probability amplitude of the
"causal action". Due to  Section 14.2 from Ref. 4: "The causal Green
function $D_{m^{2}}^{c}(x - y)$ (it seems this function was introduced
first by Rivier and Stueckelberg${}^{8}$) plays first fiddle in the
quantum field theory and describes the causal connection between the
birth and annihilation processes at the different space-time points $x$
and $y$." For the formulation of the relativistic quantum mechanics
equations describing the electromagnetic interaction we used in Ref. 6 the
distribution (\ref{2.13}). This choice allows to avoid diverging integrals.

The operator valued distribution (\ref{2.18}), (\ref{2.17}) satisfies
the relation (\ref{2.7}) for $n = 2$. Let us consider the relation (\ref{2.7})
for $n = 3$
\begin{equation}
\label{2.901} T(u_{\alpha (1)}(x_{1}) u_{\alpha (2)}(x_{2})
u_{\alpha (3)}(x_{3})) = (- 1)^{\sigma (i,j,k)} u_{\alpha
(i)}(x_{i}) u_{\alpha (j)}(x_{j}) u_{\alpha (k)}(x_{k}), \, \,
x_{i}^{0} > x_{j}^{0} > x_{k}^{0},
\end{equation}
where $i,j,k$ is a permutation of the integers $1,2,3$. The numbers $\sigma (i,j,k)$
are given by the relations (\ref{2.702}). For the left-hand side of the equality
(\ref{2.901}) we choose the operator valued distribution
$$
T(u_{\alpha (1)}(x_{1}) u_{\alpha (2)}(x_{2}) u_{\alpha (3)}(x_{3}))
= u_{\alpha (1)}(x_{1}) u_{\alpha (2)}(x_{2})u_{\alpha (3)}(x_{3}) +
$$
\begin{equation}
\label{2.902} \sum_{i\, =\, 1}^{3} (- 1)^{\sigma (1,\widehat{i}, 3,i)} <u_{\alpha (1)}(x_{1})
\widehat{u_{\alpha (i)}(x_{i})} u_{\alpha (3)}(x_{3})>_{c} u_{\alpha (i)}(x_{i})
\end{equation}
similar to the operator valued distribution in the right-hand side of the
equality (\ref{2.18}).

Any free Fermi-field commutes with any free Bose-field. It is sufficient
therefore to define the chronological product (\ref{2.902}) for the
free Fermi-fields only ($2l_{i} + 2\dot{l}_{i} = 1\, \, \hbox{mod} \, 2$,
$i = 1,2,3$) or for the free Bose-fields only
($2l_{i} + 2\dot{l}_{i} = 0\, \, \hbox{mod} \, 2$, $i = 1,2,3$). The
conditions
\begin{equation}
\label{2.903}  (2l_{1} + 2\dot{l}_{1}) (2l_{2} + 2\dot{l}_{2}) =
(2l_{1} + 2\dot{l}_{1}) (2l_{3} + 2\dot{l}_{3}) \, \, \hbox{mod} \, 2,
\end{equation}
\begin{equation}
\label{2.904}
(2l_{2} + 2\dot{l}_{2}) (2l_{1} + 2\dot{l}_{1}) =
(2l_{2} + 2\dot{l}_{2}) (2l_{3} + 2\dot{l}_{3}) \, \, \hbox{mod} \, 2,
\end{equation}
\begin{equation}
\label{2.905}  (2l_{3} + 2\dot{l}_{3}) (2l_{1} + 2\dot{l}_{1}) =
(2l_{3} + 2\dot{l}_{3}) (2l_{2} + 2\dot{l}_{2}) \, \, \hbox{mod} \, 2
\end{equation}
are fulfilled both for the free Fermi-fields and for the free Bose-fields.
The conditions (\ref{2.903}), (\ref{2.904}) imply the condition (\ref{2.905}).

In view of the relations (\ref{2.21}), (\ref{2.17}) the
distribution $<u_{\alpha (1)}(x_{1}) \widehat{u_{\alpha (i)}(x_{i})}
u_{\alpha (3)}(x_{3})>_{c} \, \,  =\, \, 0$
for $x_{1}^{0} > \widehat{x_{i}^{0}} > x_{3}^{0}$, $i = 1,2,3$. Hence
the operator valued distribution (\ref{2.902}) coincides with the right-hand
side of the relation (\ref{2.901}) in the domain $x_{1}^{0} > x_{2}^{0} > x_{3}^{0}$.
Due to the relations (\ref{2.700}), (\ref{2.702}), (\ref{2.18}), (\ref{2.21}),
(\ref{2.17}) the operator valued distribution (\ref{2.902}) restriction to the
domain $x_{2}^{0} > x_{1}^{0} > x_{3}^{0}$ coincides with the right-hand side
of the relation (\ref{2.901}). Due to the relations (\ref{2.700}), (\ref{2.702}),
(\ref{2.18}), (\ref{2.21}), (\ref{2.17}) the operator valued distribution
(\ref{2.902}) restriction to the domain $x_{1}^{0} > x_{3}^{0} > x_{2}^{0}$
coincides with the right-hand side of the relation (\ref{2.901}), if the
condition (\ref{2.903}) is valid. Due to the relations (\ref{2.700}),
(\ref{2.702}), (\ref{2.18}), (\ref{2.21}), (\ref{2.17}) the operator valued
distribution (\ref{2.902}) restriction to the domain $x_{3}^{0} > x_{1}^{0} > x_{2}^{0}$
coincides with the right-hand side of the relation (\ref{2.901}), if the
condition (\ref{2.903}) is valid. Due to the relations (\ref{2.700}),
(\ref{2.702}), (\ref{2.18}), (\ref{2.21}), (\ref{2.17}) the operator valued
distribution (\ref{2.902}) restriction to the domain $x_{2}^{0} > x_{3}^{0} > x_{1}^{0}$
coincides with the right-hand side of the relation (\ref{2.901}), if the
condition (\ref{2.904}) is valid. Due to the relations (\ref{2.700}),
(\ref{2.702}), (\ref{2.18}), (\ref{2.21}), (\ref{2.17}) the operator valued
distribution (\ref{2.902}) restriction to the domain $x_{3}^{0} > x_{2}^{0} > x_{1}^{0}$
coincides with the right-hand side of the relation (\ref{2.901}), if the
condition (\ref{2.904}) is valid.

The adjoint free spin field $\bar{\psi}_{\alpha}(x_{1})$ and the free spin field
$\psi_{\beta} (x_{2})$ are Fermi-fields. The free electromagnetic field $A_{\mu} (x_{3})$
is Bose-field and commutes with fields $\bar{\psi}_{\alpha}(x_{1})$ and $\psi_{\beta} (x_{2})$.
By making use of the operator valued distribution (\ref{2.18}) we define the chronological
electromagnetic interaction Lagrangian
\begin{equation}
\label{2.908}
\sum_{\alpha, \beta \, =\, 1}^{4} \sum_{\mu \, =\, 0}^{3}
e\gamma_{\alpha \beta}^{\mu} T(\bar{\psi}_{\alpha}(x_{1}) \psi_{\beta} (x_{2})) A_{\mu} (x_{3}).
\end{equation}
Let us define the chronological product for a free Bose field, namely for the
free electromagnetic field
$$
T(A_{\mu (1)}(x_{1})\cdots A_{\mu (n)}(x_{n}))
= A_{\mu (1)}(x_{1}) \cdots A_{\mu (n)}(x_{n}) +
$$
$$
\sum_{1\, \leq \, k\, <\, l\, \leq \, n} <A_{\mu (k)}(x_{k})A_{\mu
(l)}(x_{l})>_{c} \times
$$
\begin{equation}
\label{2.906} A_{\mu (1)}(x_{1}) \cdots
\widehat{A_{\mu (k)}(x_{k})} \cdots \widehat{A_{\mu (l)}(x_{l})}
\cdots A_{\mu (n)}(x_{n})  + \cdots
\end{equation}
The operator valued distributions (\ref{2.18}) and (\ref{2.906}) are similar.
The equality (\ref{2.906}) is assumed in "Causal Scattering Matrix and the
Chronological Product," arXiv: 1011.0859, as the definition of the chronological
product for a free quantum field. The subsequent summings in the right-hand of the
equality (\ref{2.906}) run over two pairs of numbers from $1,..,n$, run over three
pairs of numbers from $1,..,n$, etc.

Let us rewrite the definition (\ref{2.906}) in the recurrent way
$$
T(I) = I,\, \, T(A_{\mu} (x)) = A_{\mu} (x),
$$
\begin{eqnarray}
\label{2.907} T(A_{\mu (1)}(x_{1}) \cdots  A_{\mu (n + 1)}(x_{n +
1})) = T(A_{\mu (1)}(x_{1}) \cdots  A_{\mu (n)}(x_{n}))A_{\mu (n +
1)}(x_{n + 1}) + \nonumber
\\ \sum_{k\, =\, 1}^{n} <A_{\mu (k)}(x_{k})A_{\mu (n + 1)}(x_{n + 1})>_{c}
T(A_{\mu (1)}(x_{1}) \cdots  \widehat{A_{\mu (k)}(x_{k})} \cdots
A_{\mu (n)}(x_{n})).
\end{eqnarray}
We consider the permutation $1,2 \rightarrow j(1),j(2)$. The second commutation
relation (\ref{2.1}), the relation (\ref{2.21}) and the second
relation (\ref{2.17}) imply the relation (\ref{2.7}),
$n = 2$. We suppose the relation (\ref{2.7}) for the chronological product
$T(A_{\mu (1)}(x_{1})\cdots A_{\mu (m)}(x_{m}))$ for any integer $m = 2,..,n$.
Then we obtain the relation (\ref{2.7}) for $n + 1$ fields by making use of
the second commutation relation (\ref{2.1}), the relation (\ref{2.21}), the
second relation (\ref{2.17}) and the recurrent relation (\ref{2.907}).

We rewrite the relation (\ref{2.906})
$$
T(A_{\mu (1)} (x_{1})\cdots A_{\mu (n)} (x_{n})) = A_{\mu (1)}(x_{1})
\cdots A_{\mu (n)}(x_{n}) +
$$
$$
\sum_{1\, \leq \, k\, <\, l\, \leq \, n} A_{\mu (1)}(x_{1}) \cdots
A_{\mu (k - 1)}(x_{k - 1}) \times
$$
$$
(T(A_{\mu (k)} (x_{k})A_{\mu (l)} (x_{l})) - A_{\mu (k)} (x_{k})A_{\mu
(l)} (x_{l})) \times
$$
\begin{equation}
\label{3.21} A_{\mu (k + 1)}(x_{k + 1})\cdots \widehat{A_{\mu
(l)}(x_{l})} \cdots A_{\mu (n)}(x_{n}) + \cdots
\end{equation}
taking into account the relation (\ref{2.18}) for the free electromagnetic field.
The subsequent summings in the right-hand of the equality (\ref{3.21}) run over
two pairs of numbers from $1,..,n$, run over three pairs of numbers from $1,..,n$,
etc. We have placed the operator

\noindent $T(A_{\mu (k)}(x_{k})A_{\mu (l)}(x_{l})) - A_{\mu (k)}(x_{k})A_{\mu
(l)}(x_{l})$ instead of the operator $A_{\mu (k)}(x_{k})$ in the relation
(\ref{3.21}). In view of the equality (\ref{2.18}) it is possible to place
this operator instead of the operator $A_{\mu (l)}(x_{l})$.

The second relation (\ref{2.1}), the relation (\ref{2.21}) and the second
relation (\ref{2.17}) imply the relation (\ref{2.18}) for the free
electromagnetic field. The relation (\ref{2.18}) for the free electromagnetic
field implies the relation
$$
A_{\mu (1)}(x_{1}) \cdots A_{\mu (n)}(x_{n}) + \sum_{1\, \leq \, k\,
<\, j(1)} A_{\mu (1)}(x_{1}) \cdots A_{\mu (k - 1)}(x_{k - 1})
\times
$$
$$
(T(A_{\mu (k)}(x_{k})A_{\mu (j(1))}(x_{j(1)})) - A_{\mu
(k)}(x_{k})A_{\mu (j(1))}(x_{j_{1}})) \times
$$
$$
A_{\mu (k + 1)}(x_{k + 1})\cdots \widehat{A_{\mu
(j(1))}(x_{j(1)})} \cdots A_{\mu (n)}(x_{n})=
$$
\begin{equation}
\label{3.22} A_{\mu (j(1))}(x_{j(1)})A_{\mu (1)}(x_{1})\cdots
\widehat{A_{\mu (j(1))}(x_{j(1)})} \cdots A_{\mu (n)}(x_{n}),
\end{equation}
$$
x_{j(1)}^{0} > x_{j(2)}^{0} > \cdots > x_{j(n)}^{0},
$$
for the permutation $1,...,n \rightarrow j(1),...,j(n)$. (For the free
electromagnetic field

\noindent $\sigma (j(1),1,...,\widehat{j(1)},...,n) = 0\, \, \hbox{mod} \, 2$.)
If we consider now the sum over the variable $1 \leq k < l_{1} = j(1)$
and the double sum over the variables $1 \leq k_{1} < l_{1} = j(1)$,
$1 \leq k_{2} < l_{2} = j(2)$ in the left-hand side of the equality of
the type (\ref{3.22}), then we get the operator
$$
A_{\mu (j(1))}(x_{j(1)})A_{\mu (j(2))}(x_{j(2)}) A_{\mu
(1)}(x_{1})\cdots \widehat{A_{\mu (j(1))}(x_{j(1)})} \cdots
\widehat{A_{\mu (j(2))}(x_{j(2)})} \cdots A_{\mu (n)}(x_{n}).
$$
in the right-hand side of the equality of the type (\ref{3.22}).
If we continue this process, then we prove the relation (\ref{2.7})
for the free electromagnetic field. Hence the relations (\ref{2.18})
and (\ref{3.21}) imply the relation (\ref{2.7}) for the free
electromagnetic field.

The relations (\ref{2.18}), (\ref{3.21}) for the free electromagnetic field imply
$$
T(A_{\mu (1)}(x_{1}) \cdots A_{\mu (n)}(x_{n})) =
$$
\begin{equation}
\label{3.221}
T(A_{\mu (j(1))}(x_{1}) \cdots A_{\mu (j(k))}(x_{(j(k))}))
T(A_{\mu (j(k + 1))}(x_{k + 1}) \cdots A_{\mu (j(n))}(x_{(j(n))})),
\end{equation}
$$
x_{j(p)}^{0} > x_{j(q)}^{0},\, \,  p = 1,...,k,\, \, q = k + 1,...,n,
$$
$$
j(1) < \cdots < j(k),\, \, j(k + 1) < \cdots < j(n),
$$
for the permutation $1,...,n \rightarrow j(1),...,j(n)$.

Let us substitute the relations (\ref{2.23}) for the free electromagnetic field
\begin{equation}
\label{3.27} <A_{\mu (k)}(x_{k})A_{\mu (l)}(x_{l})>_{c} \, \, =\, \,
<T(A_{\mu (k)}(x_{k})A_{\mu (l)}(x_{l}))>_{0} - <A_{\mu (k)}(x_{k})
A_{\mu (l)}(x_{l})>_{0}
\end{equation}
into the chronological product definition (\ref{2.906}). Choose the
term $- <A_{\mu (k)} (x_{k}) A_{\mu (l)} (x_{l})>_{0}$ in every
sum (\ref{3.27}) of the equality (\ref{2.906}). Adding the first term
$A_{\mu (1)} (x_{1}) \cdots A_{\mu (n)} (x_{n})$ we get the first term
in the right-hand side of the equality
$$
T(A_{\mu (1)} (x_{1}) \cdots A_{\mu (n)} (x_{n})) = \, \, :A_{\mu
(1)} (x_{1}) \cdots A_{\mu (n)} (x_{n}): +
$$
$$
\sum_{1\, \leq \, k\, <\, l\, \leq \, n} <T(A_{\mu (k)} (x_{k})
A_{\mu (l)} (x_{l}))>_{0} \times
$$
\begin{equation}
\label{3.28} :A_{\mu (1)} (x_{1}) \cdots \widehat{A_{\mu (k)}
(x_{k})} \cdots \widehat{A_{\mu (l)} (x_{l})} \cdots A_{\mu (n)}
(x_{n}): + \cdots
\end{equation}
due to the relation (\ref{2.5}) for the free electromagnetic field.
Let us choose the first term $<T(A_{\mu (k)}(x_{k})A_{\mu (l)}(x_{l}))>_{0}$
in one sum (\ref{3.27}) of the relation (\ref{2.906}) and choose the
second term $ - <A_{\mu (k)} (x_{k}) A_{\mu (l)} (x_{l})>_{0}$ in all
other sums (\ref{3.27}). We get the second term in the right-hand side
of the equality (\ref{3.28}). If we continue this process, then we prove the
relation (\ref{3.28}). The subsequent summings in the right-hand of the
equality (\ref{3.28}) run over two pairs of numbers from $1,..,n$, run over
three pairs of numbers from $1,..,n$, etc. The definition (\ref{3.28}) is
similar to the Wick theorem for the chronological products (Ref. 4, Section 19.2).

\section{III. SCATTERING MATRIX}
\setcounter{equation}{0}
For the quantum electrodynamics we propose the interaction operator
\begin{equation}
\label{3.12} L_{3}(x_{1},x_{2},x_{3})
h_{3}(x_{1},x_{2},x_{3}) = \sum_{\alpha, \beta \, =\, 1}^{4} \sum_{\mu \, =\, 0}^{3}
e\gamma_{\alpha \beta}^{\mu} :\bar{\psi}_{\alpha}
(x_{1}) \psi_{\beta} (x_{2}):A_{\mu} (x_{3})
h_{3}(x_{1},x_{2},x_{3}).
\end{equation}
The smooth switching function $h_{3}(x_{1},x_{2},x_{3})$ has the compact support.
The Fermi-operators are included in the even combination:
$:\bar{\psi}_{\alpha} (x_{1})\psi_{\beta} (x_{2}):$. In order to define the
chronological product of the operators (\ref{3.12}) we have no need of
operator valued distribution of the type (\ref{2.902}). It is sufficient to
define the chronological product of  the bilocal products
$\bar{\psi}_{\alpha } (x_{1}) \psi_{\beta } (x_{2})$. The operator valued
distribution (\ref{3.12}) is called bilocal (polylocal in Section 16.8 from Ref. 4).
The word "polylocal" means "bilocal" for two fields. The integral of the interaction
operator (\ref{3.12}) may be extended to the switching distribution (\ref{3.3}).
We assume
\begin{equation}
\label{3.37}
<I>_{0}\, \, = 1,\, \,
<:u_{\alpha (1)}(x_{1}) \cdots u_{\alpha (n)}(x_{n}):>_{0}\, \, =\, \, 0,\, \, n \geq 1,
\end{equation}
and define the product of free fields
$$
(\Phi_{n(1),n(2),n(3)}, f_{n(1) + n(2) + n(3)}) =
\int d^{4}x_{1} \cdots d^{4}x_{n(1)}d^{4}y_{1}\cdots d^{4}y_{n(2)}
d^{4}z_{1} \cdots d^{4}z_{n(3)}\times
$$
$$
\bar{\psi}_{\alpha (1)} (x_{1}) \cdots
\bar{\psi}_{\alpha (n(1))} (x_{n(1)})\psi_{\beta (1)}
(y_{1}) \cdots \psi_{\beta (n(2))} (y_{n(2)})A_{\mu (1)} (z_{1}) \cdots
A_{\mu (n(3))} (z_{n(3)}) \times
$$
\begin{equation}
\label{3.40}
f_{n(1) + n(2) + n(3)}(x_{1},...,x_{n(1)},y_{1},...,y_{n(2)},z_{1},...,z_{n(3)}).
\end{equation}
$f_{n(1) + n(2) + n(3)}(x_{1},...,x_{n(1)},y_{1},...,y_{n(2)},z_{1},...,z_{n(3)})$
is a smooth function with a compact support. By making use of the relations
(\ref{2.4}), (\ref{3.12}) and (\ref{3.37}) we get the vacuum expectation
$$
<\Phi_{n(1),n(2),n(3)} (x_{1},...,x_{n(1)},y_{1},...,y_{n(2)},
z_{n(3)},...,z_{n(3)}
L_{3}(x,y,z)\times
$$
$$
\Phi_{m(1),m(2),m(3)} (x_{n(1) + 1},...,x_{n(1) + m(1)},y_{n(2) + 1},...,y_{n(2) + m(2)},
z_{n(3) + 1},...,z_{n(3) + m(3)}>_{0}\, \, =
$$
$$
\sum (- 1)^{\sigma} e\gamma_{\alpha \beta}^{\mu} <\bar{\psi}_{\alpha}
(x) \psi_{\beta (j)}
(y_{j})>_{0} <\bar{\psi}_{\alpha (k)} (x_{k})\psi_{\beta} (y)>_{0}
<A_{\mu} (z)A_{\mu (l)} (z_{l})>_{0} \times
$$
\begin{equation}
\label{3.38} \prod <\bar{\psi}_{\alpha (j(1))}
(x_{j(1)}) \psi_{\beta (j(2))} (y_{j(2)})>_{0}
<A_{\mu (k(1))} (z_{k(1)})A_{\mu (k(2))} (z_{k(2)})>_{0} +  \cdots
\end{equation}
The subsequent summings in the right-hand side of the equality (\ref{3.38})
contain the terms with another order of free fields. Due to the relation
(\ref{2.402}) the terms containing the multiplier
$<\bar{\psi}_{\alpha} (x) \psi_{\beta }(y)>_{0}$ and are absent. Hence the
distribution (\ref{3.38}) may be extended to the switching distribution (\ref{3.3}).
If we insert the distribution (\ref{3.3}) into the integral of the interaction
operator (\ref{3.12}), then we get
\begin{equation}
\label{3.13} \int d^{4}x L(x) g(x) =
\int d^{4}x \sum_{\alpha, \beta \, =\, 1}^{4} \sum_{\mu \, =\, 0}^{3}
e\gamma_{\alpha \beta}^{\mu} :\bar{\psi}_{\alpha} (x) \psi_{\beta}
(x):A_{\mu} (x)g(x)
\end{equation}
where $L(x)$ is the interaction Lagrangian (\ref{3.2}). The bilocal operator
$L(x)$ is called local in Section 16.8 from Ref. 4.

In order to define the chronological product of two interaction operators (\ref{3.12})
we define the chronological product of two bilocal products
$\bar{\psi}_{\alpha } (x) \psi_{\beta } (y)$
\begin{equation}
\label{3.15}
T(;\bar{\psi}_{\alpha (1)} (x_{1}) \psi_{\beta (1)} (y_{1});
\bar{\psi}_{\alpha (2)} (x_{2}) \psi_{\beta (2)} (y_{2});) =
\end{equation}
$$
\bar{\psi}_{\alpha (1)} (x_{1}) \psi_{\beta (1)}
(y_{1})\bar{\psi}_{\alpha (2)} (x_{2}) \psi_{\beta (2)} (y_{2})\, \, +
$$
$$
<\bar{\psi}_{\alpha (1)} (x_{1}) \psi_{\beta (2)} (y_{2})>_{c}
\psi_{\beta (1)} (y_{1})\bar{\psi}_{\alpha (2)} (x_{2})\, \, +
$$
$$
<\psi_{\beta (1)} (y_{1})\bar{\psi}_{\alpha (2)}
(x_{2})>_{c} \bar{\psi}_{\alpha (1)} (x_{1}) \psi_{\beta (2)}
(y_{2})\, \, +
$$
$$
<\bar{\psi}_{\alpha (1)} (x_{1}) \psi_{\beta (2)}
(y_{2})>_{c} <\psi_{\beta (1)} (y_{1})\bar{\psi}_{\alpha (2)}
(x_{2})>_{c}.
$$
The right-hand side of the equality (\ref{3.15}) is the operator valued
distribution everywhere defined. Due to the relations (\ref{2.1}), (\ref{2.21}),
(\ref{2.17}) this operator valued distribution is equal to
$\bar{\psi}_{\alpha (2)} (x_{2}) \psi_{\beta (2)}
(y_{2})\bar{\psi}_{\alpha (1)} (x_{1}) \psi_{\beta (1)} (y_{1})$
in the domain $\{ x_{2}^{0},y_{2}^{0}\} > \{ x_{1}^{0},y_{1}^{0}\}$ and is
equal to $\bar{\psi}_{\alpha (1)} (x_{1}) \psi_{\beta (1)}
(y_{1})\bar{\psi}_{\alpha (2)} (x_{2}) \psi_{\beta (2)} (y_{2})$ in
the domain $\{ x_{1}^{0},y_{1}^{0}\} > \{ x_{2}^{0},y_{2}^{0}\}$. The inequality

\noindent $\{ x_{1}^{0},x_{2}^{0}\} > \{ y_{1}^{0},y_{2}^{0}\}$ means four
inequalities $x_{i}^{0} > y_{j}^{0}$, $i,j = 1,2$. The chronological product of
two bilocal interaction operators (\ref{3.12}) is defined by the following relation
$$
T(;L_{3}(x_{1},y_{1},z_{1}); L_{3}(x_{2},y_{2},z_{2});) =
$$
$$
\sum_{\alpha (1), \alpha (2), \beta (1), \beta (2) \, =\, 1}^{4} \,
\, \sum_{\mu (1), \mu (2) \, =\, 0}^{3} e^{2}\gamma_{\alpha (1)
\beta (1)}^{\mu (1)} \gamma_{\alpha (2) \beta (2)}^{\mu (2)}
T(A_{\mu (1)} (z_{1}) A_{\mu (2)} (z_{2})) \times
$$
\begin{equation}
\label{3.16}
T(;:\bar{\psi}_{\alpha (1)} (x_{1}) \psi_{\beta (1)} (y_{1}):;
:\bar{\psi}_{\alpha (2)} (x_{2}) \psi_{\beta (2)} (y_{2}):;)
\end{equation}
where the chronological product of two bilocal normal products
$:\bar{\psi}_{\alpha } (x_{1}) \psi_{\beta } (x_{2}):$
$$
T(;:\bar{\psi}_{\alpha (1)} (x_{1}) \psi_{\beta (1)} (y_{1}):;
:\bar{\psi}_{\alpha (2)} (x_{2}) \psi_{\beta (2)} (y_{2}):;)\, \, =
$$
$$
T(;\bar{\psi}_{\alpha (1)} (x_{1}) \psi_{\beta (1)} (y_{1});
\bar{\psi}_{\alpha (2)} (x_{2}) \psi_{\beta (2)} (y_{2});)\, \, -
$$
$$
<\bar{\psi}_{\alpha (1)} (x_{1}) \psi_{\beta (1)} (y_{1})>_{0}
\bar{\psi}_{\alpha (2)} (x_{2}) \psi_{\beta (2)} (y_{2})\, \, -
$$
$$
<\bar{\psi}_{\alpha (2)} (x_{2}) \psi_{\beta (2)}
(y_{2})>_{0} \bar{\psi}_{\alpha (1)} (x_{1}) \psi_{\beta (1)}
(y_{1})\, \, +
$$
\begin{equation}
\label{3.160}
<\bar{\psi}_{\alpha (1)} (x_{1}) \psi_{\beta (1)}
(y_{1})>_{0} <\bar{\psi}_{\alpha (2)} (x_{2}) \psi_{\beta (2)}
(y_{2})>_{0}.
\end{equation}
The chronological product $ T(A_{\mu_{1}} (z_{1}) A_{\mu_{2}}
(z_{2}))$ is defined by the relation (\ref{2.906}). By making use the
relations (\ref{2.4}), (\ref{2.18}), (\ref{2.23}), (\ref{3.15}), (\ref{3.160})
it is possible to prove the equality
$$
T(;:\bar{\psi}_{\alpha (1)} (x_{1}) \psi_{\beta (1)} (y_{1}):;
:\bar{\psi}_{\alpha (2)} (x_{2}) \psi_{\beta (2)} (y_{2}):;) =
$$
$$
:\bar{\psi}_{\alpha (1)} (x_{1}) \psi_{\beta (1)}
(y_{1})\bar{\psi}_{\alpha (2)} (x_{2}) \psi_{\beta (2)} (y_{2}): +
$$
$$
<T(\bar{\psi}_{\alpha (1)} (x_{1}) \psi_{\beta (2)} (y_{2}))>_{0} \,
\, :\psi_{\beta (1)} (y_{1})\bar{\psi}_{\alpha (2)} (x_{2}): +
$$
$$
 <T(\psi_{\beta (1)} (y_{1})\bar{\psi}_{\alpha (2)}
(x_{2}))>_{0} \, \, :\bar{\psi}_{\alpha (1)} (x_{1})
\psi_{\beta (2)} (y_{2}): +
$$
\begin{equation}
\label{3.161}
<T(\bar{\psi}_{\alpha (1)} (x_{1}) \psi_{\beta (2)} (y_{2}))>_{0}
<T(\psi_{\beta (1)} (y_{1})\bar{\psi}_{\alpha (2)} (x_{2}))>_{0}.
\end{equation}
The scattering matrix in quantum electrodynamics is given by the
relation
$$
S_{N}(h_{3}(x_{1},x_{2},x_{3})) = I + i \int d^{4}xd^{4}yd^{4}z
L_{3}(x,y,z)h_{3}(x,y,z) +
$$
$$
\sum_{m\, =\, 2}^{N} \frac{i^{m}}{m!} \int
d^{4}x_{1}d^{4}y_{1}d^{4}z_{1}\cdots d^{4}x_{m} d^{4}y_{m}d^{4}z_{m}\times
$$
\begin{equation}
\label{3.17} T\left( ;L_{3}(x_{1},y_{1},z_{1});\cdots ;
L_{3}(x_{m},y_{m},z_{m});\right) h_{3}(x_{1},y_{1},z_{1})\cdots
h_{3}(x_{m},y_{m},z_{m}).
\end{equation}
The first term in the right-hand side of the equality (\ref{3.17}) is the
identity operator $I$. The second term is proportional to the integral of
the interaction operator (\ref{3.12}). The third term in the right-hand side
of the equality (\ref{3.17}) is
$$
 - \frac{1}{2} \int d^{4}x_{1}d^{4}y_{1}d^{4}z_{1}d^{4}x_{2}
d^{4}y_{2}d^{4}z_{2}\times
$$
\begin{equation}
\label{3.19} T\left( ;L_{3}(x_{1},y_{1},z_{1});
L_{3}(x_{2},y_{2},z_{2});\right) h_{3}(x_{1},y_{1},z_{1})
h_{3}(x_{2},y_{2},z_{2}).
\end{equation}
The chronological product
$T\left( ;L_{3}(x_{1},y_{1},z_{1});L_{3}(x_{2},y_{2},z_{2});\right)$
contains the products of the distributions (\ref{2.9}) with the different arguments
in view of the relations (\ref{3.16}), (\ref{3.161}).
The integral (\ref{3.19}) is convergent for the smooth function $h_{3}(x,y,z)$
rapidly decreasing at the infinity. The operator (\ref{3.19}) can not be extended
to the switching distribution (\ref{3.3}). If we insert the switching distribution
(\ref{3.3}) into the integral (\ref{3.19}), we get
\begin{equation}
\label{3.18} - \frac{1}{2}
\int d^{4}x_{1}d^{4}x_{2}
T\left( ;L_{3}(x_{1},x_{1},x_{1});L_{3}(x_{2},x_{2},x_{2});\right) g(x_{1})g(x_{2}).
\end{equation}
The chronological product
$T\left( ;L_{3}(x_{1},x_{1},x_{1});L_{3}(x_{2},x_{2},x_{2});\right)$ contains
the products of the distributions
$<T(\bar{\psi}_{\alpha (1)} (x_{1}) \psi_{\beta (2)} (x_{2}))>_{0}$,
$<T(\psi_{\beta (1)} (x_{1})\bar{\psi}_{\alpha (2)} (x_{2}))>_{0}$ and

\noindent $<T(A_{\mu (1)} (x_{1}) A_{\mu (2)} (x_{2}))>_{0}$ with the same argument
$x_{2} - x_{1}$. The integral (\ref{3.18}) is divergent.

For the bilocal products of free spin fields we define the chronological products
$$
T(;I:) = I,\, \,
T(;\bar{\psi}_{\alpha (1)} (x_{1}) \psi_{\alpha (2)} (x_{2});)\, \,
=\, \, \bar{\psi}_{\alpha (1)} (x_{1}) \psi_{\alpha (2)} (x_{2}),
$$
$$
T(;\bar{\psi}_{\alpha (1)} (x_{1}) \psi_{\alpha (2)} (x_{2}); \cdots;
\bar{\psi}_{\alpha (2n - 1)} (x_{2n - 1}) \psi_{\alpha (2n)} (x_{2n});)\, \,
=\, \, B_{1} \cdots B_{n} +
$$
\begin{equation}
\label{3.23}
\sum_{1\, \leq \, k\, <\, l\, \leq \, n} B_{1} \cdots B_{k - 1}
(T(;B_{k};B_{l};) - B_{k}B_{l})B_{k + 1}\cdots
\widehat{B_{l}} \cdots B_{n} + \cdots
\end{equation}
$$
B_{i} = \bar{\psi}_{\alpha (2i - 1)} (x_{2i - 1})
\psi_{\alpha (2i)} (x_{2i}),\, \, i = 1,...,n,\, \, n \geq 2.
$$
The chronological products (\ref{3.23}) are similar to the chronological
products (\ref{3.21}). The subsequent summing in the right-hand side of
the equality (\ref{3.23}) runs over two pairs of the integers from $1,...,n$
and etc. The chronological product $T(;B_{k};B_{l};)$ is given by the relation
(\ref{3.15}). For $n = 2$ the relation (\ref{3.23}) coincides with the relation
(\ref{3.15}). For the bilocal products of free spin fields the relation
$$
T(;\bar{\psi}_{\alpha (1)} (x_{1}) \psi_{\beta (1)}
(y_{1}); \cdots; \bar{\psi}_{\alpha (n)} (x_{n}) \psi_{\beta (n)}
(y_{n});)\, \, =
$$
\begin{equation}
\label{3.24} \bar{\psi}_{\alpha (j(1))} (x_{j(1)})
\psi_{\beta (j(1))} (y_{j(1)}) \cdots \bar{\psi}_{\alpha (j(n))}
(x_{j(n)}) \psi_{\beta (j(n))} (y_{j(n)}),
\end{equation}
$$
\{ x_{j(1)}^{0},y_{j(1)}^{0} \} > \{
x_{j(2)}^{0},y_{j(2)}^{0} \} > \cdots > \{ x_{j(n)}^{0},y_{j(n)}^{0} \},
$$
is valid. The relation (\ref{3.24}) proof is similar to the proof of the
relation (\ref{2.7}) for the free electromagnetic field. The first proof
uses the relation (\ref{3.23}). The last proof uses the relation (\ref{3.21}).

For the bilocal products of free spin fields the relation
$$
T(;\bar{\psi}_{\alpha (1)} (x_{1}) \psi_{\beta (1)}
(y_{1}); \cdots; \bar{\psi}_{\alpha (n)} (x_{n}) \psi_{\beta (n)}
(y_{n});)\, \, =
$$
$$
T(;\bar{\psi}_{\alpha (j(1))} (x_{j(1)})
\psi_{\beta (j(1))} (y_{j(1)}); \cdots :\bar{\psi}_{\alpha (j(k))}
(x_{j(k)}) \psi_{\beta (j(k))} (y_{j(k)});) \times
$$
\begin{equation}
\label{3.247}
T(;\bar{\psi}_{\alpha (j(k + 1))} (x_{j(k + 1)})
\psi_{\beta (j(k + 1))} (y_{j(k + 1)}); \cdots ;\bar{\psi}_{\alpha (j(n))}
(x_{j(n)}) \psi_{\beta (j(n))} (y_{j(n)});),
\end{equation}
$$
\{ x_{j(p)}^{0},y_{j(p)}^{0} \} > \{ x_{j(q)}^{0},y_{j(q)}^{0} \},\, \,
p = 1,...,k,\, \, q = k + 1,...,n,
$$
$$
j(1) < \cdots < j(k),\, \, j(k + 1) < \cdots < j(n),
$$
is valid for the permutation $1,...,n \rightarrow j(1),...,j(n)$. The relation
(\ref{3.247}) proof is similar to the proof of the relation (\ref{3.221}) for
the free electromagnetic field. The first proof uses the relation (\ref{3.23}).
The last proof uses the relation (\ref{3.21}).

The operator $T(;B_{k};B_{l};) - B_{k}B_{l}$ stands in the operator $B_{k}$
place in the definition (\ref{3.23}). The operator $B_{l}$ place is free.
Let us rewrite the definition (\ref{3.23}) in the symmetric form. For the bilocal
products the definitions (\ref{2.4}), (\ref{3.23}) and the relations (\ref{2.53}),
(\ref{2.23}) imply
\begin{equation}
\label{3.241}
T(;\bar{\psi}_{\alpha (1)} (x_{1}) \psi_{\alpha (2)} (x_{2}); \cdots;
\bar{\psi}_{\alpha (2n - 1)} (x_{2n - 1}) \psi_{\alpha (2n)} (x_{2n});)\, \,
=
\end{equation}
$$
:\bar{\psi}_{\alpha (1)} (x_{1})\psi_{\alpha (2)} (x_{2}) \cdots \bar{\psi}_{\alpha (2n - 1)}
(x_{2n - 1}) \psi_{\alpha (2n)} (x_{2n}): +
$$
$$
\sum_{1\, \leq \, 2k - 1\, <\, 2l - 1\, < \, 2n}
<T(\bar{\psi}_{\alpha (2k - 1)} (x_{2k - 1})\psi_{\alpha (2l)} (x_{2l}))>_{0} (- 1)^{\sigma (2k - 1,2l,1,...,\widehat{2k - 1},...,\widehat{2l},...,2n)} \times
$$
$$
:\bar{\psi}_{\alpha (1)} (x_{1})\psi_{\alpha (2)} (x_{2}) \cdots
\widehat{\bar{\psi}_{\alpha (2k - 1)} (x_{2k - 1})} \cdots
\widehat{\psi_{\alpha (2l)} (x_{2l})} \cdots \bar{\psi}_{\alpha (2n - 1)}
(x_{2n - 1}) \psi_{\alpha (2n)} (x_{2n}): +
$$
$$
\sum_{1\, < \, 2k\, <\, 2l\, \leq \, 2n}
<T(\psi_{\alpha (2k)} (x_{2k})\bar{\psi}_{\alpha (2l - 1)} (x_{2l - 1}))>_{0} (- 1)^{\sigma (2k,2l - 1,1,...,\widehat{2k},...,\widehat{2l - 1},...,2n)} \times
$$
$$
:\bar{\psi}_{\alpha (1)} (x_{1})\psi_{\alpha (2)} (x_{2}) \cdots
\widehat{\psi_{\alpha (2k)} (x_{2k})} \cdots
\widehat{\bar{\psi}_{\alpha (2l - 1)} (x_{2l - 1})} \cdots \bar{\psi}_{\alpha (2n - 1)}
(x_{2n - 1}) \psi_{\alpha (2n)} (x_{2n}): +
$$
$$
\sum_{1\, \leq \, 2k - 1\, < \, 2n}
<\bar{\psi}_{\alpha (2k - 1)} (x_{2k - 1})\psi_{\alpha (2k)} (x_{2k})>_{0}
(- 1)^{\sigma (2k - 1,2k ,1,...,\widehat{2k - 1},\widehat{2k},...,2n)} \times
$$
$$
:\bar{\psi}_{\alpha (1)} (x_{1})\psi_{\alpha (2)} (x_{2}) \cdots
\widehat{\bar{\psi}_{\alpha (2k - 1)} (x_{2k - 1})}
\widehat{\psi_{\alpha (2k)} (x_{2k})} \cdots \bar{\psi}_{\alpha (2n - 1)}
(x_{2n - 1}) \psi_{\alpha (2n)} (x_{2n}): + \cdots
$$
The subsequent summings in the right-hand of the equality (\ref{3.241}) run over
two pairs of numbers from $1,..,n$, run over three pairs of numbers from $1,..,n$,
etc. In the right-hand of the equality (\ref{3.241}) there is the average
$<T(\bar{\psi}_{\alpha (2k - 1)} (x_{2k - 1})\psi_{\alpha (2l)} (x_{2l}))>_{0}$
for the fields $\bar{\psi}_{\alpha (2k - 1)} (x_{2k - 1})$ and
$\psi_{\alpha (2l)} (x_{2l})$ from the different bilocal products
instead of the average

\noindent $<\bar{\psi}_{\alpha (2k - 1)} (x_{2k - 1})\psi_{\alpha (2l)} (x_{2l})>_{0}$
in the equality (\ref{2.4}). There is the average

\noindent $<\bar{\psi}_{\alpha (2k - 1)} (x_{2k - 1})\psi_{\alpha (2k)} (x_{2k})>_{0}$
for the fields $\bar{\psi}_{\alpha (2k - 1)} (x_{2k - 1})$ and $\psi_{\alpha (2k)} (x_{2k})$
from the same bilocal product in the equality (\ref{3.241}) as in the equality (\ref{2.4}).
We give the definition for the chronological product of the normal bilocal products
similar to the relation (\ref{3.161})
\begin{equation}
\label{3.243}
T(;:\bar{\psi}_{\alpha (1)} (x_{1}) \psi_{\alpha (2)} (x_{2}):; \cdots;
:\bar{\psi}_{\alpha (2n - 1)} (x_{2n - 1}) \psi_{\alpha (2n)} (x_{2n}):;)\, \,
=
\end{equation}
$$
:\bar{\psi}_{\alpha (1)} (x_{1})\psi_{\alpha (2)} (x_{2}) \cdots \bar{\psi}_{\alpha (2n - 1)}
(x_{2n - 1}) \psi_{\alpha (2n)} (x_{2n}): +
$$
$$
\sum_{1\, \leq \, 2k - 1\, <\, 2l - 1\, < \, 2n}
<T(\bar{\psi}_{\alpha (2k - 1)} (x_{2k - 1})\psi_{\alpha (2l)} (x_{2l}))>_{0} (- 1)^{\sigma (2k - 1,2l,1,...,\widehat{2k - 1},...,\widehat{2l},...,2n)} \times
$$
$$
:\bar{\psi}_{\alpha (1)} (x_{1})\psi_{\alpha (2)} (x_{2}) \cdots
\widehat{\bar{\psi}_{\alpha (2k - 1)} (x_{2k - 1})} \cdots
\widehat{\psi_{\alpha (2l)} (x_{2l})} \cdots \bar{\psi}_{\alpha (2n - 1)}
(x_{2n - 1}) \psi_{\alpha (2n)} (x_{2n}): +
$$
$$
\sum_{1\, < \, 2k\, <\, 2l\, \leq \, 2n}
<T(\psi_{\alpha (2k)} (x_{2k})\bar{\psi}_{\alpha (2l - 1)} (x_{2l - 1}))>_{0}
(- 1)^{\sigma (2k,2l - 1,1,...,\widehat{2k},...,\widehat{2l - 1},...,2n)} \times
$$
$$
:\bar{\psi}_{\alpha (1)} (x_{1})\psi_{\alpha (2)} (x_{2}) \cdots
\widehat{\psi_{\alpha (2k)} (x_{2k})} \cdots
\widehat{\bar{\psi}_{\alpha (2l - 1)} (x_{2l - 1})} \cdots \bar{\psi}_{\alpha (2n - 1)}
(x_{2n - 1}) \psi_{\alpha (2n)} (x_{2n}): + \cdots
$$
The subsequent summings in the right-hand of the equality (\ref{3.243}) run over
two pairs of numbers from $1,..,n$, run over three pairs of numbers from $1,..,n$,
etc. In the right-hand of the equality (\ref{3.243}) there is the average
$<T(\bar{\psi}_{\alpha (2k - 1)} (x_{2k - 1})\psi_{\alpha (2l)} (x_{2l}))>_{0}$
for the fields $\bar{\psi}_{\alpha (2k - 1)} (x_{2k - 1})$ and
$\psi_{\alpha (2l)} (x_{2l})$ from the different bilocal products
instead of the average
$<\bar{\psi}_{\alpha (2k - 1)} (x_{2k - 1})\psi_{\alpha (2l)} (x_{2l})>_{0}$
in the equality (\ref{2.4}). For the fields
$\bar{\psi}_{\alpha (2k - 1)} (x_{2k - 1})$ and $\psi_{\alpha (2k)} (x_{2k})$
from the same bilocal product the average equal to zero is considered instead
of the average
$<\bar{\psi}_{\alpha (2k - 1)} (x_{2k - 1})\psi_{\alpha (2k)} (x_{2k})>_{0}$
in the equality (\ref{2.4}).

For the normal bilocal products we prove the relation of the type (\ref{3.160})
\begin{equation}
\label{3.245}
T(;:\bar{\psi}_{\alpha (1)} (x_{1}) \psi_{\alpha (2)} (x_{2}):;\cdots ;
:\bar{\psi}_{\alpha (2n - 1)} (x_{2n - 1}) \psi_{\alpha (2n)} (x_{2n}):;) =
\end{equation}
$$
\sum_{l\, =\, 0}^{n} \, \, \sum_{1\, \leq 2k(1) - 1\, <\, \cdots \, < 2k(l) - 1\, <\, 2n}
(- 1)^{l} <B_{2k(1) - 1}>_{0} \cdots  <B_{2k(l) - 1}>_{0} \times
$$
$$
T(;B_{1}; \cdots ;\widehat{B_{2k(1) - 1}}; \cdots ;\widehat{B_{2k(l) - 1}}; \cdots  ;B_{2n - 1};),
$$
$$
B_{2k - 1} = \bar{\psi}_{\alpha (2k - 1)} (x_{2k - 1}) \psi_{\alpha (2k)} (x_{2k}),\, \,
k = 1,...,n.
$$
We insert the equalities (\ref{3.241}) into the equality (\ref{3.245}). In right-hand side
of the equality (\ref{3.245}) the terms containing the average
$<\bar{\psi}_{\alpha (2k(j) - 1)} (x_{2k(j) - 1}) \psi_{\alpha (2k(j))} (x_{2k(j)})>_{0}$
as a multiplier are included one time into the sum with $l = 0$ and are included one time
into the sum with $l = 1$. Hence these term are absent: $1 - 1 = 0$. The terms containing
two averages
$<\bar{\psi}_{\alpha (2k(i) - 1)} (x_{2k(i) - 1}) \psi_{\alpha (2k(i))} (x_{2k(i)})>_{0}$
and $<\bar{\psi}_{\alpha (2k(j) - 1)} (x_{2k(j) - 1}) \psi_{\alpha (2k(j))} (x_{2k(j)})>_{0}$
as the multipliers are included one time into the sum with $l = 0$ and are included two times
into the sum with $l = 1$ (one time among terms
$$
<\bar{\psi}_{\alpha (2k(i) - 1)} (x_{2k(i) - 1}) \psi_{\alpha (2k(i))} (x_{2k(i)})>_{0}
T(;B_{1}; \cdots ;\widehat{B_{2k(i) - 1}}; \cdots  ;B_{2n - 1};)
$$
and one time among terms
$$
<\bar{\psi}_{\alpha (2k(j) - 1)} (x_{2k(j) - 1}) \psi_{\alpha (2k(j))} (x_{2k(j)})>_{0}
T(;B_{1}; \cdots ;\widehat{B_{2k(j) - 1}}; \cdots  ;B_{2n - 1};)).
$$
These terms are included also one time into the sum with $l = 2$. Hence these terms are
absent: $1 - 2 + 1 = (1 - 1)^{2} = 0$. The terms containing three averages

\noindent $<\bar{\psi}_{\alpha (2k(i) - 1)} (x_{2k(i) - 1}) \psi_{\alpha (2k(i))} (x_{2k(i)})>_{0}$,
$<\bar{\psi}_{\alpha (2k(j) - 1)} (x_{2k(j) - 1}) \psi_{\alpha (2k(j))} (x_{2k(j)})>_{0}$
and

\noindent $<\bar{\psi}_{\alpha (2k(l) - 1)} (x_{2k(l) - 1}) \psi_{\alpha (2k(l))} (x_{2k(l)})>_{0}$
as the multipliers are included one time into the sum with $l = 0$, are included $3 = 3!(1!2!)^{- 1}$
times into the sum with $l = 1$, are included $3 = 3!(2!1!)^{- 1}$ times into the sum with
$l = 2$ and are included one time into the sum with $l = 3$. Hence these terms are absent:
$1 - 3!(1!2!)^{- 1} + 3!(2!1!)^{- 1} - 1 = (1 - 1)^{3} = 0$. If we continue this process,
then we prove that the terms containing any average

\noindent $<\bar{\psi}_{\alpha (2k(j) - 1)} (x_{2k(j) - 1}) \psi_{\alpha (2k(j))} (x_{2k(j)})>_{0}$
as a multiplier are absent. In the right-hand side of the equality (\ref{3.245}) the terms
without some average
$<\bar{\psi}_{\alpha (2k(j) - 1)} (x_{2k(j) - 1}) \psi_{\alpha (2k(j))} (x_{2k(j)})>_{0}$
as a multiplier are included one time into the sum with $l = 0$. The sum in the right-hand side
of the equality (\ref{3.243}) consists namely of these terms.

The relations (\ref{3.23}) - (\ref{3.247}), (\ref{3.245}) imply the relations
$$
T(;:\bar{\psi}_{\alpha (1)} (x_{1}) \psi_{\beta (1)}
(y_{1}):; \cdots; :\bar{\psi}_{\alpha (n)} (x_{n}) \psi_{\beta (n)}
(y_{n}):;)\, \, =
$$
$$
T(;:\bar{\psi}_{\alpha (j(1))} (x_{j(1)})
\psi_{\beta (j(1))} (y_{j(1)}):;) \cdots T(;:\bar{\psi}_{\alpha (j(n))}
(x_{j(n)}) \psi_{\beta (j(n))} (y_{j(n)}):;) =
$$
\begin{equation}
\label{3.246} :\bar{\psi}_{\alpha (j(1))} (x_{j(1)})
\psi_{\beta (j(1))} (y_{j(1)}): \cdots :\bar{\psi}_{\alpha (j(n))}
(x_{j(n)}) \psi_{\beta (j(n))} (y_{j(n)}):,
\end{equation}
$$
\{ x_{j(1)}^{0},y_{j(1)}^{0} \} > \{ x_{j(2)}^{0},y_{j(2)}^{0} \} > \cdots
> \{ x_{j(n)}^{0},y_{j(n)}^{0} \},
$$
$$
T(;:\bar{\psi}_{\alpha (1)} (x_{1}) \psi_{\beta (1)}
(y_{1}):; \cdots; :\bar{\psi}_{\alpha (n)} (x_{n}) \psi_{\beta (n)}
(y_{n}):;)\, \, =
$$
$$
T(;:\bar{\psi}_{\alpha (j(1))} (x_{j(1)})
\psi_{\beta (j(1))} (y_{j(1)}):; \cdots ;:\bar{\psi}_{\alpha (j(k))}
(x_{j(k)}) \psi_{\beta (j(k))} (y_{j(k)}):;) \times
$$
\begin{equation}
\label{3.248}
T(;:\bar{\psi}_{\alpha (j(k + 1))} (x_{j(k + 1)})
\psi_{\beta (j(k + 1))} (y_{j(k + 1)}):; \cdots ;:\bar{\psi}_{\alpha (j(n))}
(x_{j(n)}) \psi_{\beta (j(n))} (y_{j(n)}):;),
\end{equation}
$$
\{ x_{j(p)}^{0},y_{j(p)}^{0} \} > \{
x_{j(q)}^{0},y_{j(q)}^{0} \}, \, \, p = 1,...,k,\, \, q = k + 1,...,n,
$$
$$
j(1) < \cdots < j(k),\, \, j(k + 1) < \cdots < j(n),
$$
for the permutation $1,...,n \rightarrow j(1),...,j(n)$.

The chronological product of the bilocal Lagrangians (\ref{3.12}) is
defined by the relation similar to the relation (\ref{3.16})
$$
T(;L_{3}(x_{1},y_{1},z_{1}); \cdots;
L_{3}(x_{n},y_{n},z_{n});) =\, \, \sum_{\alpha (1),.., \alpha (n), \beta (1),..., \beta (n) \, =\,
1}^{4} \, \, \sum_{\mu (1),..., \mu (n) \, =\, 0}^{3}
$$
$$
e^{n}\gamma_{\alpha (1) \beta (1)}^{\mu (1)} \cdots
\gamma_{\alpha (n) \beta (n)}^{\mu (n)} T(A_{\mu (1)} (z_{1}) \cdots
A_{\mu (n)} (z_{n})) \times
$$
\begin{equation}
\label{3.25}
T(;:\bar{\psi}_{\alpha (1)} (x_{1}) \psi_{\beta (1)} (y_{1}):; \cdots;
:\bar{\psi}_{\alpha (n)} (x_{n}) \psi_{\beta (n)} (y_{n}):;).
\end{equation}
In view of the relations (\ref{2.907}), (\ref{3.12}), (\ref{3.23}) the
chronological product (\ref{3.25}) for $n = 1$ is
\begin{equation}
\label{3.47}
T\left( ;L_{3}(x,y,z);\right) = L_{3}(x,y,z).
\end{equation}
The relations (\ref{2.7}), (\ref{3.221}) for the free electromagnetic field and the relations
(\ref{3.246}), (\ref{3.248})  imply the relations for the chronological product
(\ref{3.25})
\begin{equation}
\label{3.14} T(;L_{3}(x_{1},y_{1},z_{1}); \cdots ;
L_{3}(x_{n},y_{n},z_{n});) = L_{3}(x_{j(1)},y_{j(1)},z_{j(1)})
\cdots L_{3}(x_{j(n)},y_{j(n)},z_{j(n)}),
\end{equation}
$$
\{ x_{j(1)}^{0},y_{j(1)}^{0},z_{j(1)}^{0} \} > \{
x_{j(2)}^{0},y_{j(2)}^{0},z_{j(2)}^{0} \} > \cdots
> \{ x_{j(n)}^{0},y_{j(n)}^{0},z_{j(n)}^{0} \},
$$
$$
T(;L_{3}(x_{1},y_{1},z_{1}); \cdots ;
L_{3}(x_{n},y_{n},z_{n});) =
$$
$$
T(L_{3}(x_{j(1)},y_{j(1)},z_{j(1)})
\cdots L_{3}(x_{j(k)},y_{j(k)},z_{j(k)})) \times
$$
\begin{equation}
\label{3.26} T(L_{3}(x_{j(k + 1)},y_{j(k + 1)},z_{j(k + 1)})
\cdots L_{3}(x_{j(n)},y_{j(n)},z_{j(n)})),
\end{equation}
$$
\{ x_{j(p)}^{0},y_{j(p)}^{0},z_{j(p)}^{0} \} > \{
x_{j(q)}^{0},y_{j(q)}^{0},z_{j(q)}^{0} \}, \, \, p = 1,...,k,\, \,
q = k + 1,...,n,
$$
$$
j(1) < \cdots < j(k),\, \, j(k + 1) < \cdots < j(n),
$$
for the permutation $1,...,n \rightarrow j(1),...,j(n)$. The inequality
$\{ x_{1}^{0},x_{2}^{0},x_{3}^{0} \} > \{ y_{1}^{0},y_{2}^{0},y_{3}^{0} \}$
means nine inequalities $x_{i} > y_{j}$, $i,j = 1,2,3$.

We substitute the relations (\ref{3.28}), (\ref{3.243}) into the equality (\ref{3.17}),
(\ref{3.25})
$$
S_{N}(h_{3}(x_{1},x_{2},x_{3})) = 1 + \sum_{m\, =\, 1}^{N} \frac{i^{m}}{m!}
\sum_{\alpha (1),.., \alpha (m), \beta (1),..., \beta (m) \, =\,
1}^{4} \, \, \sum_{\mu (1),..., \mu (m) \, =\, 0}^{3}
$$
$$
\int
d^{4}x_{1}d^{4}y_{1}d^{4}z_{1}\cdots d^{4}x_{m} d^{4}y_{m}d^{4}z_{m}
e^{m}\gamma_{\alpha (1) \beta (1)}^{\mu (1)} \cdots
\gamma_{\alpha (m) \beta (m)}^{\mu (m)} \times
$$
$$
:\bar{\psi}_{\alpha (1)} (x_{1}) \psi_{\beta (1)} (y_{1}) \cdots
\bar{\psi}_{\alpha (m)} (x_{m}) \psi_{\beta (m)} (y_{m}): \times
$$
$$
:A_{\mu (1)} (z_{1}) \cdots
A_{\mu (m)} (z_{m}): h_{3}(x_{1},y_{1},z_{1}) \cdots h_{3}(x_{m},y_{m},z_{m}) +
$$
$$
\sum_{m\, =\, 2}^{N} \frac{i^{m}}{m!}
\sum_{\alpha (1),.., \alpha (m), \beta (1),..., \beta (m) \, =\,
1}^{4} \, \, \sum_{\mu (1),..., \mu (m) \, =\, 0}^{3}
\, \, \sum_{1\, \leq \, k \, <\, l\, \leq \, m}
\int d^{4}x_{1}d^{4}y_{1}d^{4}z_{1}\cdots d^{4}x_{m} d^{4}y_{m}d^{4}z_{m}
$$
$$
e^{m} \gamma_{\alpha (1) \beta (1)}^{\mu (1)} \cdots
\gamma_{\alpha (m) \beta (m)}^{\mu (m)}
(- 1)^{\sigma (k,l,1,...,\widehat{k},...,\widehat{l},...,m)}
<T(\bar{\psi}_{\alpha (k)} (x_{k})\psi_{\beta (l)} (y_{l}))>_{0}
\times
$$
$$
:\bar{\psi}_{\alpha (1)} (x_{1})\psi_{\beta (1)} (y_{1}) \cdots
\widehat{\bar{\psi}_{\alpha (k)} (x_{k})} \cdots
\widehat{\psi_{\beta (l)} (y_{l})} \cdots \bar{\psi}_{\alpha (m)}
(x_{m}) \psi_{\beta (m)} (y_{m}): \times
$$
$$
:A_{\mu (1)} (z_{1}) \cdots A_{\mu (m)} (z_{m}):
h_{3}(x_{1},y_{1},z_{1}) \cdots h_{3}(x_{m},y_{m},z_{m})\, \, +
$$
$$
\sum_{m\, =\, 2}^{N} \frac{i^{m}}{m!}
\sum_{\alpha (1),.., \alpha (m), \beta (1),..., \beta (m) \, =\,
1}^{4} \, \, \sum_{\mu (1),..., \mu (m) \, =\, 0}^{3}
\, \, \sum_{1\, \leq \, k\, <\, l\, \leq \, m}
\int d^{4}x_{1}d^{4}y_{1}d^{4}z_{1}\cdots d^{4}x_{m} d^{4}y_{m}d^{4}z_{m}
$$
$$
e^{m} \gamma_{\alpha (1) \beta (1)}^{\mu (1)} \cdots
\gamma_{\alpha (m) \beta (m)}^{\mu (m)}
(- 1)^{\sigma (k,l,1,...,\widehat{k},...,\widehat{l},...,2n)}
<T(\psi_{\beta (k)} (y_{k})\bar{\psi}_{\alpha (l)} (x_{l}))>_{0}
\times
$$
$$
:\bar{\psi}_{\alpha (1)} (x_{1})\psi_{\beta (1)} (y_{1}) \cdots
\widehat{\psi_{\beta (k)} (y_{k})} \cdots
\widehat{\bar{\psi}_{\alpha (l)} (x_{l })} \cdots \bar{\psi}_{\alpha (m)}
(x_{m}) \psi_{\beta (m)} (y_{m}): \times
$$
$$
:A_{\mu (1)} (z_{1}) \cdots A_{\mu (m)} (z_{m}):
h_{3}(x_{1},y_{1},z_{1}) \cdots h_{3}(x_{m},y_{m},z_{m})\, \, +
$$
$$
\sum_{m\, =\, 2}^{N} \frac{i^{m}}{m!}
\sum_{\alpha (1),.., \alpha (m), \beta (1),..., \beta (m) \, =\,
1}^{4} \, \, \sum_{\mu (1),..., \mu (m) \, =\, 0}^{3}
\, \, \sum_{1\, \leq \, k\, <\, l\, \leq \, m} \int
d^{4}x_{1}d^{4}y_{1}d^{4}z_{1}\cdots d^{4}x_{m} d^{4}y_{m}d^{4}z_{m}
$$
$$
h_{3}(x_{1},y_{1},z_{1}) \cdots h_{3}(x_{m},y_{m},z_{m})
e^{m}\gamma_{\alpha (1) \beta (1)}^{\mu (1)} \cdots
\gamma_{\alpha (m) \beta (m)}^{\mu (m)} \times
$$
$$
<T(A_{\mu (k)} (z_{k})
A_{\mu (l)} (z_{l}))>_{0} :\bar{\psi}_{\alpha (1)} (x_{1}) \psi_{\beta (1)} (y_{1}) \cdots
\bar{\psi}_{\alpha (m)} (x_{m}) \psi_{\beta (m)} (y_{m}): \times
$$
\begin{equation}
\label{3.39}
:A_{\mu (1)} (z_{1}) \cdots \widehat{A_{\mu (k)}
(z_{k})} \cdots \widehat{A_{\mu (l)} (z_{l})} \cdots A_{\mu (n)}
(z_{m}): \, \, + \cdots
\end{equation}
The subsequent sum consists of the terms containing the distributions (\ref{2.9}) products.

Due to the relation (\ref{2.21}) the support of the Lorentz invariant distribution
(\ref{2.13}) lies in the closed upper light cone. For any number $\Delta > 0$ and
for any smooth function $\phi (x)$ with a compact support for the variable $x^{0}$
the following relation and inequality from Ref. 6
$$
\int d^{4}x D_{m^{2}}^{ret}(x) \phi (x)\, \, =\, \,
-\, (2\pi)^{- 4} \int d^{4}p ((p^{0} + i\Delta )^{2} - |{\bf p}|^{2} - m^{2})^{- 1}
\times
$$
\begin{equation}
\label{3.32}
\int dx^{0} \exp \{ (\Delta  - ip^{0})x^{0}\} \tilde{\phi} (x^{0},{\bf p}),\, \,
\tilde{\phi} (x^{0},{\bf p}) = \int d^{3}{\bf x}
\exp \Biggl\{ i\sum_{k\, \, =\, 1}^{3} p^{k}x^{k}\Biggr\} \phi (x^{0},{\bf x}),
\end{equation}
\begin{equation}
\label{3.34}
\Bigl|(p^{0} + i\Delta )^{2} - |{\bf p}|^{2} - m^{2}\Bigr|^{- 1}
\leq \Delta^{- 2} (1 + (p^{0})^{2})^{- 1}(1 + |{\bf p}|^{2} + m^{2} + 2\Delta^{2})
\end{equation}
imply the inequality
\begin{equation}
\label{3.33}
\Biggl| \int d^{4}x D_{m^{2}}^{ret}(x) \phi (x) \Biggr| \, \, \leq \, \,
C(\Delta)   \sup_{{\bf p} \, \in \, {\bf R}^{3}}
(1 + |{\bf p}|^{2} + m^{2} + 2\Delta^{2})^{3} \int_{0}^{\infty} dx^{0} \exp \{ \Delta x^{0}\}
|\tilde{\phi} (x^{0},{\bf p})|.
\end{equation}
The relations (\ref{2.17}) and the inequality (\ref{3.33}) imply the inequalities
$$
\Biggl| \int d^{4}x  d^{4}y  <A_{\mu} (x) A_{\nu} (y)>_{c} f(x) g(y)
\Biggr| \, \, \leq
$$
\begin{equation}
\label{3.330}
C(\Delta)   \sup_{{\bf p} \, \in \, {\bf R}^{3}}
(1 + |{\bf p}|^{2} + 2\Delta^{2})^{3} \int_{0}^{\infty} dx^{0} \exp \{ \Delta x^{0}\}
|\tilde{\phi}_{\mu \nu} (x^{0},{\bf p})|,
\end{equation}
$$
\phi_{\mu \nu} (x) = i\eta^{\mu \nu} \int d^{4}y g(x + y)f(y),
$$
$$
\Biggl| \int d^{4}x  d^{4}y <\psi_{\alpha} (x)\bar{\psi_{\beta}} (y)>_{c} f(x) g(y)
\Biggr| \, \, \leq
$$
\begin{equation}
\label{3.331}
C(\Delta)   \sup_{{\bf p} \, \in \, {\bf R}^{3}}
(1 + |{\bf p}|^{2} + m^{2} + 2\Delta^{2})^{3} \int_{0}^{\infty} dx^{0} \exp \{ \Delta x^{0}\}
|\tilde{\phi}_{\alpha \beta} (x^{0},{\bf p})|,
\end{equation}
$$
\phi_{\alpha \beta} (x) = - \int d^{4}y \left( \sum_{\mu \, =\, 0}^{3} \gamma_{\alpha
\beta}^{\mu} \frac{\partial}{\partial x^{\mu}} + im\right) g(x + y)f(y).
$$
For the distribution $D_{m^{2}}^{-}(x)$ defined by the relation (\ref{3.8}) and
for a smooth function $\phi (x)$ rapidly decreasing at the infinity the inequality
\begin{equation}
\label{3.35}
\Biggl| \int d^{4}x D_{m^{2}}^{-}(x) \phi (x) \Biggr| \, \, \leq
C\,  \sup_{{\bf p} \, \in \, {\bf R}^{3}}
\, \, (1 + |{\bf p}|^{2})^{2} \int dx^{0} |\tilde{\phi} (x^{0}, {\bf p})|
\end{equation}
is valid. The relations (\ref{2.3}) and the inequality (\ref{3.35}) imply the inequalities
\begin{equation}
\label{3.351} \Biggl| \int d^{4}x  d^{4}y <A_{\mu}(x)A_{\nu}(y)>_{0} f(x) g(y)
\Biggr| \, \, \leq \, \,
C\,  \sup_{{\bf p} \, \in \, {\bf R}^{3}}
\, \, (1 + |{\bf p}|^{2})^{2} \int dx^{0} |\tilde{\phi}_{\mu \nu} (x^{0}, {\bf p})|,
\end{equation}
$$
\phi_{\mu \nu} (x) = i\eta^{\mu \nu} \int d^{4}y f(x + y)g(y),
$$
\begin{equation}
\label{3.352} \Biggl| \int d^{4}x  d^{4}y <\psi_{\alpha} (x)\bar{\psi}_{\beta} (y)>_{0} f(x) g(y)
\Biggr| \, \, \leq \, \,
C\,  \sup_{{\bf p} \, \in \, {\bf R}^{3}}
\, \, (1 + |{\bf p}|^{2})^{2} \int dx^{0} |\tilde{\phi}_{\alpha \beta} (x^{0}, {\bf p})|,
\end{equation}
$$
\phi_{\alpha \beta} (x) = -  \left(
\sum_{\mu \, =\, 0}^{3} \gamma_{\alpha \beta}^{\mu}
\frac{\partial}{\partial x^{\mu}} + im\right) \int d^{4}y f(x + y)g(y).
$$
We estimate the first sum in the equality (\ref{3.39}). In view of the equalities
(\ref{2.4}), (\ref{3.37})
$$
<(\Phi_{n(1),n(2),n(3)}, f_{n(1) + n(2) + n(3)}) \Biggl( \frac{i^{m}}{m!}
\sum_{\alpha (1),.., \alpha (m), \beta (1),..., \beta (m) \, =\,
1}^{4} \, \, \sum_{\mu (1),..., \mu (m) \, =\, 0}^{3}
$$
$$
\int
d^{4}x_{1}d^{4}y_{1}d^{4}z_{1}\cdots d^{4}x_{m} d^{4}y_{m}d^{4}z_{m}
e^{m}\gamma_{\alpha (1) \beta (1)}^{\mu (1)} \cdots
\gamma_{\alpha (m) \beta (m)}^{\mu (m)} \times
$$
$$
:\bar{\psi}_{\alpha (1)} (x_{1}) \psi_{\beta (1)} (y_{1}) \cdots
\bar{\psi}_{\alpha (m)} (x_{m}) \psi_{\beta (m)} (y_{m}): :A_{\mu (1)} (z_{1}) \cdots
A_{\mu (m)} (z_{m}): \times
$$
\begin{equation}
\label{3.49}
h_{3}(x_{1},y_{1},z_{1}) \cdots h_{3}(x_{m},y_{m},z_{m})
\Biggr) (\Phi_{m(1),m(2),m(3)}, g_{m(1) + m(2) + m(3)})>_{0} = 0,
\end{equation}
if $m > n(j) + m(j)$ for some $j = 1,2,3$. Due to the inequalities (\ref{3.351}),
(\ref{3.352}) the average
$$
<(\Phi_{n(1),n(2),n(3)}, f_{n(1) + n(2) + n(3)}) \Biggl( \sum_{m\, =\, 2}^{\infty} \frac{i^{m}}{m!}
\sum_{\alpha (1),.., \alpha (m), \beta (1),..., \beta (m) \, =\,
1}^{4} \, \, \sum_{\mu (1),..., \mu (m) \, =\, 0}^{3}
$$
$$
\int
d^{4}x_{1}d^{4}y_{1}d^{4}z_{1}\cdots d^{4}x_{m} d^{4}y_{m}d^{4}z_{m}
e^{m}\gamma_{\alpha (1) \beta (1)}^{\mu (1)} \cdots
\gamma_{\alpha (m) \beta (m)}^{\mu (m)} \times
$$
$$
:\bar{\psi}_{\alpha (1)} (x_{1}) \psi_{\beta (1)} (y_{1}) \cdots
\bar{\psi}_{\alpha (m)} (x_{m}) \psi_{\beta (m)} (y_{m}): :A_{\mu (1)} (z_{1}) \cdots
A_{\mu (m)} (z_{m}): \times
$$
\begin{equation}
\label{3.41}
h_{3}(x_{1},y_{1},z_{1}) \cdots h_{3}(x_{m},y_{m},z_{m})
\Biggr) (\Phi_{m(1),m(2),m(3)}, g_{m(1) + m(2) + m(3)})>_{0}
\end{equation}
is finite. In view of the inequalities (\ref{3.330}), (\ref{3.331}) the smooth switching
function $h_{3}(x,y,z)$ with a compact support may be chosen in such a way that the averages
of the second, third and fourth sums in the right-hand side of the equality (\ref{3.39})
for $N = \infty$
$$
<(\Phi_{n(1),n(2),n(3)}, f_{n(1) + n(2) + n(3)}) \Biggl(
\sum_{m\, =\, 2}^{\infty} \frac{i^{m}}{m!}
\sum_{\alpha (1),.., \alpha (m), \beta (1),..., \beta (m) \, =\,
1}^{4} \, \, \sum_{\mu (1),..., \mu (m) \, =\, 0}^{3}
\, \, \sum_{1\, \leq \, k \, <\, l\, \leq \, m}
$$
$$
\int d^{4}x_{1}d^{4}y_{1}d^{4}z_{1}\cdots d^{4}x_{m} d^{4}y_{m}d^{4}z_{m}
e^{m} \gamma_{\alpha (1) \beta (1)}^{\mu (1)} \cdots
\gamma_{\alpha (m) \beta (m)}^{\mu (m)} \times
$$
$$
(- 1)^{\sigma (k,l,1,...,\widehat{k},...,\widehat{l},...,m)}
<T(\bar{\psi}_{\alpha (k)} (x_{k})\psi_{\beta (l)} (y_{l}))>_{0}
\times
$$
$$
:\bar{\psi}_{\alpha (1)} (x_{1})\psi_{\beta (1)} (y_{1}) \cdots
\widehat{\bar{\psi}_{\alpha (k)} (x_{k})} \cdots
\widehat{\psi_{\beta (l)} (y_{l})} \cdots \bar{\psi}_{\alpha (m)}
(x_{m}) \psi_{\beta (m)} (y_{m}): \times
$$
$$
:A_{\mu (1)} (z_{1}) \cdots A_{\mu (m)} (z_{m}):
h_{3}(x_{1},y_{1},z_{1}) \cdots h_{3}(x_{m},y_{m},z_{m})\Biggr) \times
$$
$$
(\Phi_{m(1),m(2),m(3)}, g_{m(1) + m(2) + m(3)})>_{0},
$$
$$
<(\Phi_{n(1),n(2),n(3)}, f_{n(1) + n(2) + n(3)}) \Biggl(
\sum_{m\, =\, 2}^{\infty} \frac{i^{m}}{m!}
\sum_{\alpha (1),.., \alpha (m), \beta (1),..., \beta (m) \, =\,
1}^{4} \, \, \sum_{\mu (1),..., \mu (m) \, =\, 0}^{3}
\, \, \sum_{1\, \leq \, k\, <\, l\, \leq \, m}
$$
$$
\int d^{4}x_{1}d^{4}y_{1}d^{4}z_{1}\cdots d^{4}x_{m} d^{4}y_{m}d^{4}z_{m}
e^{m} \gamma_{\alpha (1) \beta (1)}^{\mu (1)} \cdots
\gamma_{\alpha (m) \beta (m)}^{\mu (m)} \times
$$
$$
(- 1)^{\sigma (k,l,1,...,\widehat{k},...,\widehat{l},...,2n)}
<T(\psi_{\beta (k)} (y_{k})\bar{\psi}_{\alpha (l)} (x_{l}))>_{0}
\times
$$
$$
:\bar{\psi}_{\alpha (1)} (x_{1})\psi_{\beta (1)} (y_{1}) \cdots
\widehat{\psi_{\beta (k)} (y_{k})} \cdots
\widehat{\bar{\psi}_{\alpha (l)} (x_{l })} \cdots \bar{\psi}_{\alpha (m)}
(x_{m}) \psi_{\beta (m)} (y_{m}): \times
$$
$$
:A_{\mu (1)} (z_{1}) \cdots A_{\mu (m)} (z_{m}):
h_{3}(x_{1},y_{1},z_{1}) \cdots h_{3}(x_{m},y_{m},z_{m})\Biggr) \times
$$
$$
(\Phi_{m(1),m(2),m(3)}, g_{m(1) + m(2) + m(3)})>_{0},
$$
$$
<(\Phi_{n(1),n(2),n(3)}, f_{n(1) + n(2) + n(3)}) \Biggl(
\sum_{m\, =\, 2}^{\infty} \frac{i^{m}}{m!}
\sum_{\alpha (1),.., \alpha (m), \beta (1),..., \beta (m) \, =\,
1}^{4} \, \, \sum_{\mu (1),..., \mu (m) \, =\, 0}^{3}
\, \, \sum_{1\, \leq \, k\, <\, l\, \leq \, m}
$$
$$
\int d^{4}x_{1}d^{4}y_{1}d^{4}z_{1}\cdots d^{4}x_{m} d^{4}y_{m}d^{4}z_{m}
e^{m} \gamma_{\alpha (1) \beta (1)}^{\mu (1)} \cdots
\gamma_{\alpha (m) \beta (m)}^{\mu (m)} \times
$$
$$
<T(A_{\mu (k)} (z_{k})
A_{\mu (l)} (z_{l}))>_{0} :\bar{\psi}_{\alpha (1)} (x_{1}) \psi_{\beta (1)} (y_{1}) \cdots
\bar{\psi}_{\alpha (m)} (x_{m}) \psi_{\beta (m)} (y_{m}): \times
$$
$$
:A_{\mu (1)} (z_{1}) \cdots \widehat{A_{\mu (k)}
(z_{k})} \cdots \widehat{A_{\mu (l)} (z_{l})} \cdots A_{\mu (n)}
(z_{m}):  \times
$$
\begin{equation}
\label{3.44}
h_{3}(x_{1},y_{1},z_{1}) \cdots h_{3}(x_{m},y_{m},z_{m})
\Biggr) (\Phi_{m(1),m(2),m(3)}, g_{m(1) + m(2) + m(3)})>_{0}
\end{equation}
are small. Quite similarly the average of the operator (\ref{3.39}) for $N = \infty$
\begin{equation}
\label{3.61}
<(\Phi_{n(1),n(2),n(3)}, f_{n(1) + n(2) + n(3)})
S_{\infty}(h_{3}(x_{1},x_{2},x_{3})) (\Phi_{m(1),m(2),m(3)}, g_{m(1) + m(2) + m(3)})>_{0}
\end{equation}
is finite for the specially chosen switching function $h_{3}(x,y,z)$ with the
compact support.

The every term in the sums (\ref{3.41}), (\ref{3.44}) is proportional at least
to two switching functions $h_{3}(x,y,z)$. In view of the inequalities (\ref{3.330}),
(\ref{3.331}), (\ref{3.351}), (\ref{3.352}) the smooth switching function $h_{3}(x,y,z)$
may be chosen in such a way that the norms (\ref{3.351}), (\ref{3.352}) are small and
$$
<(\Phi_{n(1),n(2),n(3)}, f_{n(1) + n(2) + n(3)}) \Biggl( S_{\infty}(h_{3}(x_{1},x_{2},x_{3})) - I -
i\int d^{4}xd^{4}yd^{4}z \times
$$
\begin{equation}
\label{3.45}
L_{3}(x,y,z)h_{3}(x,y,z) \Biggr) (\Phi_{m(1),m(2),m(3)}, g_{m(1) + m(2) + m(3)})>_{0}
\, \, \approx \, \, 0.
\end{equation}
The relation (\ref{3.45}) is the average of relation similar to (\ref{3.5}) for the
Lagrangian (\ref{3.2}).

The definitions (\ref{3.17}), (\ref{3.25}) imply the relation
$$
<(\Phi_{n(1),n(2),n(3)}, f_{n(1) + n(2) + n(3)}) \Biggl(
S_{\infty}(h_{3}((\Lambda (A))^{- 1} x_{1},(\Lambda (A))^{- 1} x_{2},
(\Lambda (A))^{- 1} x_{3}))
$$
\begin{equation}
\label{3.30}
- U(A)S_{\infty}(h_{3}(x_{1},x_{2},x_{3}))(U(A))^{- 1}
\Biggr) (\Phi_{m(1),m(2),m(3)}, g_{m(1) + m(2) + m(3)})>_{0} \, \, =\, \, 0.
\end{equation}
The relation (\ref{3.30}) is the average of relation similar to (\ref{3.4}).

Let us consider the scattering matrix (\ref{3.17}) with the switching function
\begin{equation}
\label{3.50}
h_{3}(x_{1},x_{2},x_{3}) =
h_{3}^{(1)}(x_{1},x_{2},x_{3}) + h_{3}^{(2)}(x_{1},x_{2},x_{3}).
\end{equation}
The function $h_{3}^{(i)}(x_{1},x_{2},x_{3})$ support lies in the bounded
domain $G_{i}^{\times 3}$, $i = 1,2$. All time points of the domain $G_{2}$ lie in
the future relative to all time points of the domain $G_{1}$. We chose the coefficients
$i^{m}(m!)^{- 1}$ in the relation (\ref{3.39}). If the coefficients in the relation
(\ref{3.39}) are $a^{m}(m!)^{- 1}$, then for any complex number $a$ the relation
(\ref{3.26}) implies the relation
$$
<(\Phi_{n(1),n(2),n(3)}, f_{n(1) + n(2) + n(3)}) \Biggl(
S_{\infty}(h_{3}^{(1)}(x_{1},x_{2},x_{3}) + h_{3}^{(2)}(x_{1},x_{2},x_{3})) \, \, -
$$
\begin{equation}
\label{3.31}
S_{\infty}(h_{3}(x_{1},x_{2},x_{3})^{(2)})S_{\infty}(h_{3}^{(1)}(x_{1},x_{2},x_{3}))
\Biggr) (\Phi_{m(1),m(2),m(3)}, g_{m(1) + m(2) + m(3)})>_{0} \, \, =\, \, 0.
\end{equation}
The relation (\ref{3.31}) is the average of relation similar to (\ref{3.9}).

The initial state norm $||\Phi ( - \infty)||$ is equal to the final state norm
$||\Phi (\infty)||$ in the non-relativistic quantum mechanics. The scattering matrix
is a unitary operator. The scattering matrix of the relativistic quantum electrodynamics
is defined by the relation (\ref{3.17}). For the real switching function $h_{3}(x,y,z)$
$$
S_{\infty}(h_{3}(x_{1},x_{2},x_{3}))S_{\infty}^{\ast}(h_{3}(x_{1},x_{2},x_{3})) = I + \int d^{4}x_{1}d^{4}y_{1}d^{4}z_{1}
h_{3}(x_{1},y_{1},z_{1})\times
$$
$$
i\left( T\left( ;L_{3}(x_{1},y_{1},z_{1});\right)  -
\left( T\left( ;L_{3}(x_{1},y_{1},z_{1});\right)
\right)^{\ast} \right) +
$$
$$
\sum_{k\, =\, 2}^{\infty}
\int d^{4}x_{1}d^{4}y_{1}d^{4}z_{1}\cdots d^{4}x_{k} d^{4}y_{k}d^{4}z_{k}
h_{3}(x_{1},y_{1},z_{1})\cdots h_{3}(x_{k},y_{k},z_{k})\times
$$
$$
\frac{i^{k}}{k!} \Biggl( T\left( ;L_{3}(x_{1},y_{1},z_{1});\cdots ;
L_{3}(x_{k},y_{k},z_{k});\right)  +
$$
$$
\sum_{m\, =\, 1}^{k\, -\, 1}
\frac{(- 1)^{m}k!}{(k - m)!m!} T\left( ;L_{3}(x_{1},y_{1},z_{1});\cdots ;
L_{3}(x_{k - m},y_{k - m},z_{k - m});\right) \times
$$
$$
\left( T\left( ;L_{3}(x_{k - m + 1},y_{k - m + 1},z_{k - m + 1});\cdots ;
L_{3}(x_{k},y_{k},z_{k});\right) \right)^{\ast} +
$$
\begin{equation}
\label{3.46}
(- 1)^{k}  \left( T\left( ;L_{3}(x_{1},y_{1},z_{1});\cdots ;
L_{3}(x_{k},y_{k},z_{k});\right) \right)^{\ast} \Biggr).
\end{equation}
The chronological product
$
T\left( ;L_{3}(x_{1},y_{1},z_{1}); \cdots ;
L_{3}(x_{n},y_{n},z_{n});\right)
$
satisfies the relation (\ref{3.14}). The adjoint operator
$
\left( T\left( ;L_{3}(x_{1},y_{1},z_{1}); \cdots ;
L_{3}(x_{n},y_{n},z_{n});\right) \right)^{\ast}
$
is called the antichronological product. Due to the second relation
(\ref{3.2}) and the definitions (\ref{2.402}), (\ref{3.12}) we have
\begin{equation}
\label{3.54} L_{3}(x,y,z) =
\sum_{\alpha, \beta \, =\, 1}^{4} \sum_{\mu \, =\, 0}^{3}
e(\gamma^{0} \gamma^{\mu})_{\alpha \beta}
\left( (\psi_{\alpha} (x))^{\ast} \psi_{\beta} (y) -
<(\psi_{\alpha} (x))^{\ast} \psi_{\beta} (y)>_{0} \right) A_{\mu} (z).
\end{equation}
The matrix $\gamma^{0 } \gamma^{0}$ is the identity $4\times 4$ - matrix
(see Ref. 4, definition (6.18)). In view of the second relation (\ref{3.2})
and of the relation (\ref{3.55}) we have
\begin{equation}
\label{3.52} <(\psi_{\alpha} (x))^{\ast}\psi_{\beta} (y))>_{0} \, \, =\, \,
\left( \sum_{\mu \, =\, 0}^{3} (\gamma^{\mu} \gamma^{0})_{\beta \alpha}
\frac{\partial}{\partial x^{\mu}} + im\gamma_{\beta \alpha}^{0} \right)
D_{m^{2}}^{-}(x - y).
\end{equation}
The relations (\ref{3.8}), (\ref{2.13}) imply
\begin{equation}
\label{3.48} \overline{ D_{m^{2}}^{-}(x)} \, \, =\, \,  - \, D_{m^{2}}^{-}(- x),
\, \,  \overline{ D_{m^{2}}^{ret}(x)} \, \, =\, \,  D_{m^{2}}^{ret}(x).
\end{equation}
The matrices $\gamma^{\mu} \gamma^{0} = 2\eta^{\mu 0} - \gamma^{0} \gamma^{\mu}$
and $\gamma^{0}$ are Hermitian (see Ref. 4, definition (6.18)). Then
the relation (\ref{3.52}) and the first relation (\ref{3.48}) imply
\begin{equation}
\label{3.53} \overline{<(\psi_{\alpha} (x))^{\ast}\psi_{\beta} (y)>_{0}} \, \, =
\, \, (\Omega, ((\psi_{\alpha} (x))^{\ast}\psi_{\beta} (y))^{\ast} \Omega)\, \, =\, \,
(\Omega, (\psi_{\beta} (y))^{\ast}\psi_{\alpha} (x) \Omega).
\end{equation}
We assume
\begin{equation}
\label{3.58}
((\psi_{\alpha} (x))^{\ast}\psi_{\beta} (y))^{\ast}
= (\psi_{\beta} (y))^{\ast}\psi_{\alpha} (x)
\end{equation}
in accordance with the relation (\ref{3.53}). The second relation (\ref{2.3}) and
the first relation (\ref{3.48}) imply
\begin{equation}
\label{3.56} \overline{<A_{\mu}(x)A_{\nu}(y)>_{0}} \, \, = \, \,
(\Omega, (A_{\mu}(x)A_{\nu}(y))^{\ast} \Omega) =\, \, (\Omega, A_{\nu}(y)A_{\mu}(x) \Omega).
\end{equation}
We assume
\begin{equation}
\label{3.59}
(A_{\mu}(x))^{\ast} = A_{\mu}(x)
\end{equation}
in accordance with the relation (\ref{3.56}). The relations (\ref{3.47}), (\ref{3.54}),
(\ref{3.53}), (\ref{3.58}) and (\ref{3.59}) imply
\begin{equation}
\label{3.51} (T(; L_{3}(x,y,z);))^{\ast} \, \, =\, \,
( L_{3}(x,y,z))^{\ast} \, \, =\, \,  L_{3}(y,x,z).
\end{equation}
The unitary condition (Ref. 4, relation (17.22)): the right-hand side of the
equality (\ref{3.46}) for the switching distribution (\ref{3.3}) is equal to
the identity operator $I$. The first term of the right-hand side of the
equality (\ref{3.46}) is equal to the identity operator $I$. The second term
of the right-hand side of the equality (\ref{3.46}) is equal to zero for the
real symmetric,
\begin{equation}
\label{3.60}
h_{3}(x_{2},x_{1},x_{3}) = h_{3}(x_{1},x_{2},x_{3}),
\end{equation}
switching function due to the relations (\ref{3.47}), (\ref{3.51}). It means
that the integral of the interaction operator (\ref{3.12}) is a self-adjoint
operator for the real symmetric switching function. The second term of the
right-hand side of the equality (\ref{3.46}) may be extended to the switching
distribution (\ref{3.3}). The switching distribution (\ref{3.3}) satisfies the
symmetry property (\ref{3.60}). The relations (\ref{2.906}), (\ref{3.15}) -
(\ref{3.160}) imply
$$
T(;L_{3}(x_{1},y_{1},z_{1}); L_{3}(x_{2},y_{2},z_{2});) =
L_{3}(x_{1},y_{1},z_{1})L_{3}(x_{2},y_{2},z_{2}) \, \, +
$$
$$
\sum_{\alpha (1), \alpha (2), \beta (1), \beta (2) \, =\, 1}^{4} \,
\, \sum_{\mu (1), \mu (2) \, =\, 0}^{3} e^{2}\gamma_{\alpha (1)
\beta (1)}^{\mu (1)} \gamma_{\alpha (2) \beta (2)}^{\mu (2)}
\Biggl( <A_{\mu (1)} (z_{1}) A_{\mu (2)} (z_{2})>_{c} \times
$$
$$
T(;:\bar{\psi}_{\alpha (1)} (x_{1}) \psi_{\beta (1)} (y_{1}):;
:\bar{\psi}_{\alpha (2)} (x_{2}) \psi_{\beta (2)} (y_{2}):;)
\, \, +
$$
$$
<\bar{\psi}_{\alpha (1)} (x_{1}) \psi_{\beta (2)} (y_{2})>_{c}
\psi_{\beta (1)} (y_{1})\bar{\psi}_{\alpha (2)} (x_{2})
A_{\mu (1)} (z_{1}) A_{\mu (2)} (z_{2})\, \, +
$$
$$
<\psi_{\beta (1)} (y_{1})\bar{\psi}_{\alpha (2)}
(x_{2})>_{c} \bar{\psi}_{\alpha (1)} (x_{1}) \psi_{\beta (2)}
(y_{2}) A_{\mu (1)} (z_{1}) A_{\mu (2)} (z_{2})\, \, +
$$
\begin{equation}
\label{3.61} <\bar{\psi}_{\alpha (1)} (x_{1}) \psi_{\beta (2)}
(y_{2})>_{c} <\psi_{\beta (1)} (y_{1})\bar{\psi}_{\alpha (2)}
(x_{2})>_{c} A_{\mu (1)} (z_{1}) A_{\mu (2)} (z_{2}) \Biggr).
\end{equation}
If the distributions (\ref{2.17}) are equal to zero and therefore
$T(;L_{3}(x_{1},y_{1},z_{1}); L_{3}(x_{2},y_{2},z_{2});) =$
$L_{3}(x_{1},y_{1},z_{1})L_{3}(x_{2},y_{2},z_{2})$,
then the third term ($k = 2$) of the right-hand side of the equality (\ref{3.46})
is equal to zero for the real symmetric switching function
$h_{3}(x_{1},x_{2},x_{3})$.

\vskip 0,5cm

\noindent {\bf ACKNOWLEDGMENTS}

\vskip 0,5cm

This work was supported in part by the Program for Supporting
Leading Scientific Schools (Grant No. 4612.2012.1) and the RAS Program "Fundamental
Problems of Nonlinear Mechanics."

\vskip 0,5cm

\noindent ${}^{1}$Stueckelberg, E. C. G. et Rivier, D., "Causalit\'e et
structure de la Matrice $S$," Helv. Phys. Acta {\bf 23}, 215 - 222 (1950).

\noindent ${}^{2}$Bogoliubov, N.N., "Causality Condition in the Quantum Field
Theory" (in Russian), Izvestyia AN SSSR, Ser. Phys. {\bf 19}, 237 - 246
(1955).

\noindent ${}^{3}$Poincar\'e, H., "Sur la dynamique de l'\'electron,"
Rendiconti Circolo Mat. Palermo {\bf 21}, 129 - 176 (1906).

\noindent ${}^{4}$Bogoliubov, N.N. and Shirkov, D.V., {\it Introduction to
the theory of quantized fields} (Interscience, New York, 1980).

\noindent ${}^{5}$Feynman, R.P., {\it QED The Strange Theory of Light and Matter.}
(Princeton University Press, Princeton, NJ, 1985).

\noindent ${}^{6}$Zinoviev, Yu.M., "Causal electromagnetic interaction equations," J. Math. Phys.
{\bf 52}, 022302 (2011).

\noindent ${}^{7}$Vladimirov, V.S., {\it Methods of Theory of Many Complex Variables} (MIT
Press, Cambridge, MA, 1966).

\noindent ${}^{8}$Rivier, D. and Stueckelberg, E. C. G., "A Convergent
Expression for the Magnetic Moment of the Neutron," Phys. Rev.
{\bf 74}, 218 - 218 (1948).

\end{document}